\documentclass[pra,aps,showpacs]{revtex4}

\usepackage{graphicx}
\usepackage{exscale}
\usepackage{subfigure}

\newcommand{\ket}[1]{\left| #1 \right\rangle}
\newcommand{\bra}[1]{\left\langle #1 \right|}

\newcommand{\be}{\begin{equation}}
\newcommand{\ee}{\end{equation}}
\newcommand{\bea}{\begin{eqnarray}}
\newcommand{\eea}{\end{eqnarray}}

\begin{document}

\title{Non-Markovianity during the quantum Zeno effect }

\author{ A. Thilagam} 
%\email{thilaphys@gmail.com}
\address{Information Technology, Engineering and Environment, 
Mawson Institute,
University of South Australia, Australia
 5095} 
\date{\today}
\begin{abstract}
We examine the Zeno and anti-Zeno  effects in the context of non-Markovian
dynamics in  entangled spin-boson systems in contact with noninteracting reservoirs.
We identify enhanced non-Markovian signatures in 
specific two-qubit partitions of a Bell-like initial  state,  with results
showing  that the intra-qubit Zeno effect or anti-Zeno effect occurs in conjunction
with inter-qubit non-Markovian dynamics  for a range of system parameters.
The time domain of effective 
Zeno or anti-Zeno dynamics is about the same order of magnitude as the non-Markovian time scale  
of the reservoir correlation dynamics, and 
changes  in decay rate due to the Zeno mechanism appears coordinated with
information flow between specific  two-qubit partitions. 
We extend our analysis to examine  the Zeno mechanism-non-Markovianity link
using  the tripartite states arising from 
a donor-acceptor-sink model of
photosynthetic biosystems.
\end{abstract}
\maketitle

\section{Introduction}\label{c1a}

Quantum  dynamical semigroups \cite{sudar1}  operating via 
 trace preserving, completely positive (CP) 
maps \cite{choi,Kraus,neilson} with inbuilt   divisibility ~\cite{wolf,bre,sus},  are well known 
to provide tractability when characterizing the  evolution dynamics of   memoryless quantum systems. 
These semigroups  constitute the non commutative counterpart of classical Markov  theories \cite{alikil},
 and  are based  on contractions of density operators embedded  either in the
Hilbert space of Hilbert-Schmidt operators \cite{sudar1} or the 
  Banach space of trace-class operators \cite{davis,lind}, 
and involve the  forward  composition law.
Recently, studies of the
non-Markovian dynamics of evolution in open quantum systems
\cite{bre,sus,raja,ushthil} have attracted increased attention  as 
the  Markovian  model  breaks down  in the strong system-environment coupling regime or when
un-factorized initial conditions exist between the system and environment \cite{sudar2,per,smir,daj,ban}, and
for which a statistical interpretation of the density matrix is not guaranteed.
 In these instances, there is violation of complete positivity of the reduced dynamics of the 
quantum system depending on  mathematical mappings known as the  assignment 
 or extension maps \cite{per,alick,per2}. These maps provide description as 
to how a subsystem  is embedded within a larger system,
and can  assume negative values  for some correlated states. The quantum system may 
then evolve with signatures of non-Markovian dynamics with novel features of decoherence
and non-invariant states even when an equilibrium point is reached. The non-invariant states
here refer  to states which occupy  a subspace that is in a state of transience, arising from the
presence of dissipative elements.

Due to the varying characteristics of different distance measures (e.g. trace distance, Bures distance,
 Hilbert-Schmidt distance \cite{neilson,karol}),
there is no single quantifiable measure, or a unique definition of non-Markovianity in
quantum systems. In general, measures of non-Markovianity are based on  deviations from the 
continuous, memoryless, completely positive semi-group feature of Markovian evolution.
Both the concepts of divisibility\cite{wolf,sus}
and distinguishability \cite{bre} have been used  to define non-Markovianity,
and in particular the  trace distance \cite{bre}, $D[\rho_1,\rho_2]$
 is a well accepted metric measure of distinguishability of two quantum states $\rho_{1},\ \rho_2$,
that  can be  used to check the violation of 
the complete positivity during  the time-evolution  of a quantum system. 
During the   time-evolution of a quantum state under the trace-preserving CP
 map $\rho(0)\mapsto\rho(t)=\Phi(t,0)\rho(0)$,
the trace-distance  does not increase with time and 
 $D[\rho_1(t),\rho_2(t)] < D[\rho_1(0),\rho_2(0)]$.
 Such a monotonically  contractive feature
is typical of a divisible Markovian mapping on the operator space, 
and the  distinguishability between any states  is bounded by its
value at the initial state.   An increase of trace distance during 
a  time intervals is thus taken as a signature (sufficient  but not necessary) of the 
attributes of non-Markovianity.

One area in which non-Markovian signatures  has potential applications
is in the quantum Zeno effect \cite{It}, which describes the retarded time evolution 
of a quantum state subjected to frequent measurements. 
In the limiting case of infinitely frequent observations, a scenario that is  physically unattainable
on the basis of the Heisenberg uncertainty \cite{naka}, further decay is inhibited and the time evolution 
of the state comes to a standstill. The opposite
effect which leads to enhancement in time evolution is known 
as anti-Zeno effect  \cite{ob}. 
The theoretical formulations of the Zeno effect, which arises naturally from the 
 quadratic short-time behavior of quantum
evolution  have been advanced considerably by Misra and Sudarshan  \cite{Misra}
and Facchi and Pascazio \cite{FacJ,fama,facRev}. 
The concept that a measurement process  can decompose the total Hilbert space into partitions of
 orthogonal quantum Zeno subspaces  has been
examined   via the adiabatic theorem \cite{FacJ,fama}, 
as an extension of the Misra and Sudarshan theorem \cite{Misra}.
Under this scheme, the initial state is confined to  single-dimensional invariant subspace, and 
its survival probability remains unchanged over a period of time.
Accordingly, different outcomes  are eliminated and the system evolves within
the same subspaces confined by the 
total Hilbert space. However it is also possible as in Schwinger's  
non-selective measurements \cite{schw},
 where projections onto multidimensional subspaces are involved, that a
 quantum system may be  steered  away from its initial state 
 \cite{facpoint}.

In this paper, we  examine the Zeno dynamics of
quantum systems  from a quantum information perspective, by
analyzing the behavior of   dynamical semigroups during
occurrence of Zeno and anti-Zeno effects.  
There are motivations to pursue this line of investigation due to the central role played by the  decay process 
during  Zeno/anti-Zeno  effect. It is known that in unstable systems, the occurrence of 
both effects  depend on critical parameters 
such as measurement frequencies and environmental noise \cite{env}.
Several quantum systems   exhibit a  combination of Zeno and anti-Zeno effects \cite{thilzeno}
such as the nanomechanical oscillator \cite{nosci}, disordered
spin systems \cite{japko} and  localized  atomic systems \cite{rai}.
The reversal in decay,  which is apparent when the system  shifts
from displaying the  anti-Zeno effect to the Zeno effect, necessitates  some degree of reversibility  
during quantum evolutions. It is currently not clear as to the precise
nature of the entities involved during these exchanges, however 
such reversals may be linked to  feedback mechanisms within the open quantum 
system-reservoir model. This has implications for 
 the semigroup law and   contractions in the Hilbert space of 
Hilbert-Schmidt operators \cite{sudar1},  with likely  violations of forward time translations.

 While the evolution within  the projected (Zeno) subspace may be  unitary,
the overall dynamics appears complex  due to 
non-Markovian dynamics. The latter feature may be exploited to explore
 potential applications in quantum control \cite{doh,bei} and   information
processing \cite{feed} involving periodic measurements, which
 can produce either a Zeno effect or an anti-Zeno effect, 
depending on the parameters of the system.  
The possibility of protecting quantum states via
the Zeno effect has  been shown  in earlier works \cite{fra,mani,oli}.
In this work, we  highlight the appearance of 
non-Markovian signatures  during the Zeno and anti-Zeno dynamics in specific two-qubit partitions
of the spin-boson-reservoir system, with a broader aim of demonstrating
 that the context of non-Markovianity
 can reveal an alternative phenomenological perspective of the quantum Zeno
effect. 

Due to advancement in femptosecond based spectroscopy, there is currently great interest in the
quantum processes which underpins quantum coherences and
the exceptionally high efficiencies of energy transfer
observed in light-harvesting systems \cite{flem,engel,olb,caru,thilherm}. 
Accordingly, we  examine the occurrence of an 
environment induced quantum Zeno effect associated with
 dissipative photosynthetic sinks. The  sinks
form an integral component of  the reaction centers (RC) in light-harvesting systems \cite{flem}
by acting as regions where  chemical energy is generated. In this study, we 
redefine the role of photosynthetic sinks 
  as detectors of the excitation that is propagating through the system,
and  examine the impact of the Zeno effect on energy transfer mechanisms,.
Such effects  are not thoroughly explored in biochemical
systems, and the physical insight obtained from  such studies is expected to
provide useful information related to the quantum control 
of energy propagation process in artificial light-harvesting systems.

This  paper is organized as follows. In Section \ref{basic},
we present several basic elements involved in the dynamics of open quantum systems
on qualitative terms, with emphasis on the role of assignment maps in determining the nature (completely positive,
positve and negative) of dynamical maps.  A brief description  of the 
 pathological elements in density matrix operators
due to initial system-reservoir interactions is also provided in Section \ref{basic}.
 In Section \ref{zenomar} we present
a  review of Sudarshan and Misra \cite{Misra} formulations of the Zeno 
dynamics of quantum systems. 
In Section \ref{crit}, we introduce a criteria to detect non-Markovianity in Zeno dynamics,
and which will be used to analyze information flow in the spin-boson-reservoir system (described
in  Section \ref{spinbo}).
 Numerical details of the inter-qubit non-Markovianity during intra-qubit Zeno/anti-Zeno effect 
are provided in Section \ref{joint}. Using a redefined non-Markovianity measure for a tripartite
state, the Zeno mechanism  in  the photosynthetic reaction center (RC) is examined in Section\ref{lhsec}.
We present our main results and conclusions  in Section \ref{con}.

\section{Dynamics of open quantum systems and assignment maps}\label{basic}

In the standard approach involving  open quantum systems to model dissipative and  decoherence processes with
 memoryless time evolution, it is convenient to represent
 quantum states via normalized density matrices, $\rho$, which  are  hermitian ($\rho = \rho^\dagger$), and  positive definite. The positive definite attribute ensures
that density matrices possess eigenvalues that are non-negative $(0 \le \lambda_i \le 1)$ and
sum to unity (${\rm Tr}(\rho)$=1) in the complex vector space of the  Hilbert space, ${\cal H}$. 
Density operators have the advantage of being able to provide
information on the fraction of ensemble systems which exist in a given state, when 
a large number of systems are under study. If $\rho$ is representative of a pure ensemble,
then $\rho^2=\rho$, and in the case of a mixed ensemble, we obtain $0 < {\rm Tr}(\rho^2) <1$. 
To be  identifiable in experimental
setups, density matrices have to satisfy the completely positivity
condition in which positive states are mapped into positive states. 
For states  which can be expressed  in a factorized density matrix form
of a tensor product state with environment  at initial time $t$=0, a
dynamical evolution to  the  final state occurs in a tractable form.  Here 
the link between initial and final state of the system can be specified
by a semigroup of completely positive dynamical
maps that ensure that the state operators retain a  probabilistic interpretation.
The possibility of occurrence of  a collection of (not uniquely) Kraus operators \cite{Kraus},
 is only guaranteed by completely positive maps so that the associated physical process 
can still continue to be described by  CP maps
\be\label{kr}
 \rho'=\Lambda(\rho)=\sum_a K_a\rho K^\dagger_a.
\ee
where the Kraus operators $K_a$ satisfy the completeness relation,
$\sum_a K^{\dagger}_a K_a=\openone$, and the  map is trace preserving.
Using the properties of a semigroup of completely positive dynamical maps.
a quantum Markovian master equation of the Lindblad form, and which
describes the time evolution 
of the reduced open system states has been obtained as \cite{lind,gor} :
\begin{eqnarray}
\frac{d}{dt}
\rho(t)&=& -{i}[{\cal H},\rho(t)]+\sum_{k,l=1}^{d}{\cal L}_{kl}(\rho(t))
\label{gks}\\
&=&-{i}[{\cal H},\rho(t)]+\frac{1}{2} \sum_{k,l=1}^{d}
a_{kl}\left(2 \chi_k\rho(t) \chi_l^\dag- \{\chi_k^\dag \chi_l,\,\rho(t)\}\right),
\label{seq}
\\  &=& - i [{\cal H},\rho(t)] +  \sum_{k=1}^{d} {\gamma_k}
\left( {\cal L}_k\rho(t) {\cal L}_k^\dag-\frac{1}{2} \{{\cal L}_k^\dag {\cal L}_k,\,\rho(t)\}\right), 
\label{lind}
\end{eqnarray}
where the Lindblad operators, ${\cal L}_{kl}$ ( Eq.~(\ref{gks})) generate the map from the initial to the final density
operators of $\rho$.  ${\cal H}$ arises from a combination of the isolated system Hamiltonian,
${\cal H}_s$ and
a system-environment interaction operator. 
$\{\chi_k\}^{d}_{k=0}$ form the  basis in  the linear  operator space,  
with $\chi_0={\cal I}$ denoting the identity. The terms $(a_{kl})$  in Eq.~(\ref{seq})
 constitute the positive definite $d$-dimensional Hermitian 
Gorini-Kossakowski-Sudarshan matrix ${\cal A}$  \cite{gor}, with a spectrum 
characterized by the decay terms, $\{\gamma_k\}$.
The first term in Eq.~(\ref{seq}) (or (\ref{lind})) represents reversibility in system dynamics,
while the symmetrized  Lindblad operators are denoted by ${\cal L}_k$, 
in which both an operator and its hermitian conjugate are labeled by $k$.
The latter incorporate
environmental effects within the Born-Markov approximation and therefore
act as the source of  non-unitary dynamics.

While the  Lindblad form in Eq.~(\ref{lind}) ensures the positivity
of density operators at any time, it is stringent in being
applicable only
to weak system-reservoir coupling when  Markov approximation
holds, greatly simplifying the mathematical structure of the mapping procedure.
Eq.~(\ref{lind}) is therefore   not representative of
situations when complete positivity of density operators is not a necessary 
feature. There are positive maps where {\it not all} eigenvalues of the mapped density operators
are positive, and  positive states continue to be  mapped into positive states. And there
are non-positive maps in which the positive states are mapped into negative states with
at least one negative eigenvalue. As pointed out in the Introduction, the latter
may arise when the system and its environment are initially correlated.
Pechukas \cite{per} first raised the possibility  that  non-Markovianity 
may result as an artifact of the product of the initial conditions,  $\rho_s(0) \otimes \rho_r(0)$, 
where $\rho_s$ ($\rho_r$) denote the density operator specific to the system, $s$ (reservoir, $r$).
 In this regard, the  Agarwal-Redfield (AR) equation of motion is seen to violate
simple positivity in the  semigroup form of Lindblad
 for certain initial states \cite{agarwal,kohen}.

The Agarwal bath \cite{agarwal} was also noted to produce spurious
  effects in the form of density matrix negativity
 for a range of initial conditions at low temperatures \cite{talkner}.
Such violations may occur in the initial stage of evolution dynamics 
due to backflow of information from the reservoir bath at short times
comparable to the bath memory time \cite{munro}. The Lindblad form fails in these
situations and it may be appropriate to use a non-Lindblad set of relations 
to describe the system dynamics  during the initial period of quantum evolution,
in which the finite time-scale of the vibrational environment  becomes relevant.
In this regard, the non-perturbative hierarchical equations of motion (HEOM) technique \cite{tani,kreis}
which  incorporates higher vibrational energies in the bath
as well as a  finite time scale of the dynamics in the vibrational 
environment  may provide a suitable alternative, as it 
is dependent on a second-order cumulant expansion that is exact for a harmonic
bath.

There has been much  debate on the occurrence of the non-positivity attribute
in  dynamical maps \cite{per,alick,per2,sudar2,shaji}, with justifications provided \cite{per2,shaji}
to show that maps need not necessarily be in a completely positive form
to describe the evolution of open quantum systems. The physical interpretation \cite{per2,shaji}
of non-positive maps  has been   linked to mathematical mappings known as assignment maps \cite{per}.
However there still seem to be a view that any  physical evolution of reduced density 
operator {\it} MUST preserve positivity, and a 
violation of the complete positivity is considered to be unphysical,
for which quantum trajectories lack  a physical interpretation.
These views are consistent  with Alicki's \cite{alick} stand
that ``the complete positivity of the reduced dynamics should be and can be preserved."
However this reasoning has been refuted by Pechukas \cite{per2} on the basis of improper use
of ``assignment maps" outside the weak coupling regime in Ref.\cite{alick}.
Non-positivity  features are of relevance in 
experimental techniques involving quantum process tomography
in systems that possess initial correlations with the environment.
The dependence of positivity of the map on the 
interplay between the assignment map and the system-environment coupling has been recently shown \cite{modiR},
with  conditions for positivity noted to arise from  correlated system-environment states.   
Using the concept of assignment maps, an earlier work \cite{shaji} examined
the physical consistency of  states for which the dynamical map is not positive.
Such states were seen to be  not amenable to the partial trace operation
of extended systems which include specific correlations. The term ``compatibility domain"
was introduced \cite{shaji} to describe the set of states for which non-positivity is physically valid.
In forthcoming sections, we use  numerical results to show the occurrence
of non-Markovian dynamics in two model systems which incorporate
Zeno dynamics. The time domains involved during effective 
Zeno or anti-Zeno dynamics are reasonably matched with the non-Markovian time scale  
of the reservoir correlation dynamics. In the next Section we provide the basics
of the Zeno effect.

\section{Zeno dynamics and Non-Markovianity}\label{zenomar}
The Kraus operators in Eq.(\ref{kr}) may be taken  as
measurement operators, and  generalized measurements may be  represented  by a set of linear maps, where the action
of each individual map acting on the density matrix may
 not preserve its trace \cite{Sud85}. 
Here we adopt Sudarshan and Misra \cite{Misra} formulations
 of the Zeno dynamics of quantum systems, in which
measurement  in the Hilbert space ${\cal H}$  are implemented
via a   Von Neumann projection operator ${\cal P}$  (which may be one- or higher dimensional) 
with the Hilbert space range, ${\cal H}_{\cal P}$.  An initial density matrix $\rho_0$ of system $S$  
in ${\cal H}_{\cal P}={\cal P H}$  satisfies 
$\rho_0 = {\cal P} \rho_0 {\cal P} ,\; \; \mathrm{Tr} [ \rho_0 {\cal P} ] = 1$.
In the absence of any measurement, the state evolves as  \cite{Misra,fama}
\bea
\label{fe}
\rho (t) & = & U(t) \rho_0 U^\dagger (t) \\ \nonumber
U(t) & = & \exp(- i {\cal H} ^\star t)
\eea
where  ${\cal H}^\star $ is  a time-independent Hamiltonian  with a lower bound.
The probability that the
system remains within  ${\cal H}_{\cal P}$ is given by 
$P(t) = \mathrm{Tr} \left( U(t) \rho_0 U^\dagger(t) {\cal P}\right)$.
In the event of measurement at time $\tau$, the density matrix $\rho(\tau)$
transforms as $\rho(\tau) = \frac{1}{p(\tau)}\;{\cal P} U(\tau) \rho_0 U^\dagger(\tau){\cal P}$,
where the survival probability  within the subspace of ${\cal H}_{\cal P}$  appear as \cite{Misra,fama}
\begin{equation}
\label{pro1}
p(\tau) = \mathrm{Tr} \left(V(\tau) \rho_0 V^\dagger(\tau) \right)
\end{equation} 
and  $V(\tau) \equiv {\cal P} U(\tau){\cal P}$.
In the case of measurements  taken at time intervals $\tau=t/N$,  the 
survival probability of the quantum state, $\rho(t/N)$
after $N$ measurements appear as (c.f Eq.(\ref{pro1}))  as
\be
\label{survorig}
p(t/N) = \mathrm{Tr} \left( V(t/N)^N \rho_0{[V(t/N)]^N}^\dagger \right)
\ee
At very large  $N$,  transitions   outside ${\cal H}_{\cal P}$  are prohibited,
and $p(\frac{t}{N}) \rightarrow 1$, so that the monitored system persists
in the original state giving rise to the Zeno effect. 
 Misra and Sudarshan \cite{Misra} 
showed the semigroup properties of the  operator  $V(t) \equiv {\cal P} U(t){\cal P}$
at $N \rightarrow \infty$, at real time, $t$.

The projection-operator partitioning method introduced by Feshbach \cite{fesh},
divides the total Hilbert space of  ${\cal H}$  
 into two orthogonal subspaces generated by  
${\cal P}$ and its complementary projection operator ${\cal Q}=1-{\cal P}$.
We note, following Eq.(\ref{survorig}), that there exists a  probability $q(t/N)=(1 - p(t/N))$
 that the quantum system has made a transition outside  ${\cal H}_{\cal P}$ to
its ortho-complement, ${\cal H}_{\cal Q}$.
We consider that the subspace of ${\cal H}_{\cal P}$ is  spanned by ($\ket{1}_p, \ket{0}_p$),
while that of  ${\cal H}_{\cal Q}$  is spanned by ($\ket{1}_q, \ket{0}_q$). Accordingly,
 $\ket{1}_p$ denotes the ``survived" state due to measurement, 
while $\ket{1}_q$ is the ``un-survived" state. The
 measurement related decoherence due to
leakages between subspaces  may be   associated with quantum jump operators
involved  in the transfer of states  from ${\cal H}_{\cal P}  \rightarrow  {\cal H}_{\cal Q}$.
The action of these random jump operators  form the basis
of the quantum trajectory approach \cite{traj1},
where the  density operator is derived from an ensemble average
of a range of conditioned operators \cite{traj1} at the select time $t$.

We consider  an initial  state which starts from  ${\cal H}_{\cal P}$ at $t=$0, which is 
applicable to the composite state, $\ket{\psi_t}= \ket{1}_p \otimes\ket{0}_q$.
After $N$ measurements during a time duration, $t$, 
the system evolves according to the quantum trajectory dynamics model \cite{traj1}
\be
\label{evolve}
    \ket{\psi_t} = \sqrt{p(t/N)} \ket{1}_p \ket{0}_q + \sqrt{q(t/N)}\ket{0}_p \ket{1}_q,
\ee
where $\sqrt{p(t/N)}$ ($\sqrt{q(t/N)}$) is the probability amplitude associated with the state
residing in the subspace generated by ${\cal P}$ (${\cal Q}$). To simplify the analysis, 
we assume that the concurrent occupation of states in the 
two subspaces is not viable during the evolution in Eq.(\ref{evolve}).
While    a   trace preserving map is inherent in  the unitary time-evolution 
 between measurements in Eq.~(\ref{evolve}),
 non-Markovian dynamics may arise when
leakages  between  different subspaces are taken into account. This possibility
is an important consideration in this work,  and  in the analysis
of the  quantum Zeno effect in the context of non-Markovian
 evolution  dynamics.

\section{Criteria  to detect non-Markovianity  in Zeno/anti-Zeno  dynamics}\label{crit}
Under a completely positive map, there occurs no improvement in the 
  distinguishability of a set of states. Therefore a
completely positive, trace preserving dynamical map  of a Markovian
evolution on the operator space,  $\rho(0)\mapsto\rho(t)=\Phi(t,0)\rho(0)$, is monotonically  contractive with respect 
to the trace-distance,  $D[\rho_1,\rho_2]=\frac{1}{2}||\rho_1-\rho_2||$.
Here $||A||=Tr[\sqrt{A^\dagger A}]$  between   two states $\rho_{1},\ \rho_2$.
The relation
\be
\label{cri1} 
D[\rho_1(t_f),\rho_2(t_f)] < D[\rho_1(t_i),\rho_2(t_i)], \; \mbox{for} \quad t_f > t_i
\ee
is useful in identifying violation of the  monotonically  contractive  
characteristic of  divisible Markovian mapping on the operator space.
We note that Eq.(\ref{cri1}) may also imply, with the incorporation of an intermediate time $t_m$,
\be
\label{cri2} 
D[\rho_1(t_f),\rho_2(t_f)] < D[\rho_1(t_i),\rho_2(t_m)], \; \mbox{for} \quad t_f > t_i, t_f > t_m
\ee
$t_m$ may not necessarily be the same as $t_i$, and Eq.(\ref{cri2}) still applies to a
 completely positive, trace preserving dynamical map.

To establish a criteria for detecting inter-qubit non-Markovian dynamics,
we  consider two initial states which are subjected to measurements 
of varying time duration. The  trace-distance between these two states 
is then monitored after increasing the measurement duration of both states by the same amount.
To simplify the numerical analysis, and without loss in generality, we choose two initial states 
$\rho_1(t_i)=\rho_1(0), \; \rho_2(t_i)=\rho_2(\tau_1)$,  one of which is subjected to measurements of 
the ideal time duration of $\tau \rightarrow 0$, and  is therefore stagnant, while the other initial state is measured 
in intervals of  $\tau_1$ (which can be taken as variable). For the final states, we select   
$\rho_1(t_f)=\rho_1(\tau_2), \; \rho_2(t_f)=\rho_2(\tau_1+\tau_2)$,  which differ from their respective  initial
states by being subjected to measurements with  interval duration that is increased by $\tau_2$. The specifics of this model is 
convenient for analysis as will be shown in Section \ref{spinbo}. In order to determine the occurrence of 
non-Markovian dynamics, we employ a  difference function dependent on trace-distances
\begin{equation}
\label{trm} 
\Delta(\tau_1, \tau_2)= D[\rho_1(\tau_2),\rho_2(\tau_1+\tau_2)] - D[\rho_1(0),\rho_2(\tau_1)],
\end{equation} 
Positive  values of $\Delta(\tau_1, \tau_2)$  marks  the presence of  non-Markovian processes due to underlying
information flow from the  environment back to the system, however we note the different context in which 
``non-Markovian" is used here (reliant on violation of the complete positivity attribute)
as compared to a similar  term adopted by the  chemical physical community.
In the latter case, the information flow from the environment back to the system, which is absent  in the Born-Markov
quantum master equation (QME), may appear within the Markovian quantum master equation by eliminating the
Born approximation. 

It is also important to note subtle  differences between the terms, 
``violation of the complete positivity" and ``non-Markovianity",
as the occurrences of  violations of a  completely positive 
map serve only as a sufficient  but not necessary signature of non-Markovianity.
Moreover, non-violation of the positivity may not even indicate the presence
of a Markovian evolution. This highlights the complex links between dynamical maps
and non-Markovian dynamics.
 We next describe the spin-boson system
which is used as a framework to demonstrate non-Markovian features of memory effects based on Eq.(\ref{trm}),
and to seek differences between the Zeno and anti-Zeno quantum dynamics.

 \section{Zeno dynamics in the spin-boson system}\label{spinbo}
The spin-boson Hamiltonian provides an exemplary dissipative
model to examine interaction between a two-state subsystem  and a thermal reservoir 
\cite{Weiss}. Here we consider the density matrix of  a spin-boson system
associated with the Liouville equation $\frac{\partial \rho}
 {\partial t} = - i [\widehat  H_{\rm T},\rho(t)]$,
where the total Hamiltonian $\widehat  H_{\rm T} =\widehat  H_{\rm qb}+
\widehat H_{\rm os} + \widehat  H_{\rm qb-os}$ and $\widehat  H_{\rm qb}$
of the two-level qubit assumes the form $\widehat  H_{\rm qb}
= \hbar \left(\frac{\Delta \Omega}{2}\, \sigma_{z} +\Delta \sigma_{x}\right)$.
The Pauli matrices are expressed in terms of the two possible
states $(\ket{0},\ket{1})$, $\sigma_{x} = \ket{0} \bra{1}
+ \ket{1} \bra{0}$ and $\sigma_{z} = \ket{1} \bra{1}
- \ket{0} \bra{0}$. $\Delta \Omega$ is the biasing energy
while  $\Delta$  is the tunneling  amplitude. 

Each spin qubit is   coupled to 
independent reservoirs  of harmonic oscillators,
$\widehat  H_{\rm os} = \sum_{\bf q} \hbar \omega_{\bf q}\,
b_{\bf q}^{\dagger}\,b_{\bf q}$. $b_{\bf q}^{\dagger}\,$ and $b_{\bf q}\,$
are the respective  creation and  annihilation 
operators of the quantum oscillator with wave vector ${\bf q}$. 
The qubit-oscillator interaction Hamiltonian is 
linear in terms of oscillator  creation and annihilation operators
$\widehat  H_{\rm qb-os} = 
\sum_{{\bf q}}   \lambda_{_{\bf q}}\, \left ( b_{\bf q}^\dagger +
b_{\bf q}\right ) \sigma_{z}$.
The term $\lambda_{_{\bf q}}$ denotes the coupling between the qubit and the
environment and is characterized by the spectral density function,
$J(\omega)$, which we assume to be of the low-frequency form
\begin{equation}
J(\omega )=\sum_{\bf q}\lambda_{_{\bf q}}^2\delta(\omega-\omega_{\bf q})=\frac{2 \Lambda
\omega }{\omega ^{2}+\alpha^{2}},  
\label{E2}
\end{equation}
The coupling strength between the qubit and
the environment, $\Lambda$, is a product of the 
reorganization energy and $\alpha$, the characteristic frequency of
the bath. 
 $\alpha$ is  related to the reservoir correlation
time, $\alpha=\frac{1}{t_b}$, and we assume that $\alpha$ 
is small compared to the qubit energy difference,
$\Delta \Omega$. With the choice of energy unit, $\Delta \Omega$=1,
$\frac{\Lambda}{\Delta \Omega^2}$  yields a measure
of  a coupling strength that is dimensionless.

We consider an initial state of the qubit with  its corresponding 
reservoir in the vacuum state at equilibrium,
$|\phi _{i}\rangle =|1\rangle _{\mathrm{s}}\otimes
\prod_{k=1}^{N'}|0_{k}\rangle _{\mathrm{r}}=
|1\rangle _{\mathrm{s}}\otimes
\ket{{\bf 0}}_{\mathrm{r}}$ 
where $\ket{1}_{\mathrm{s}}$ ($\ket{0}_{\mathrm{s}}$) is the
upper (lower) level of the qubit, and 
 $\ket{{\bf 0}}_{\mathrm{r}}$ implies that all $N'$ wavevector modes 
of the reservoir are unoccupied in the initial state. 
 During a measurement,   a transition of a qubit from  its excited state
 to ground state is triggered,  which we consider to follow  the  simple
 mode of evolution   as in  Eq.(\ref{evolve}),
$|\phi _{i}\rangle$ proceeds in time as
\begin{equation}
|\phi _{i}\rangle \longrightarrow
 u(t) \; \ket{1}_{\mathrm{s}}
\ket{{\bf 0}}_{\mathrm{r}} + v(t) \; \ket{0}_{\mathrm{s}}
\ket{{\bf 1}}_{\mathrm{r}} ,  
\label{fstate}
\end{equation}
Thus  we consider an evolution in which the excitation is  present in one
of the two subspaces as in Eq.(\ref{evolve}).
In order to keep the problem tractable we
consider that $\ket{{\bf 1}}_{\mathrm{r}}$ denotes
a  collective state of  the reservoir,  
$|{\bf 1}\rangle _{\mathrm{r}}=\frac{1}{v(t)}
\sum_{n} \lambda _{\{n\}}(t)|\{n\}\rangle$ 
where $\{n\}$ denotes an occupation scheme in which there are
$n_i$ oscillators with wavevector $k=i$ in the reservoir
and we define the state $|\{n\}\rangle$ as $
|\{n\}\rangle =|n_0,n_1,n_2...n_i..n_{N'}\rangle$.
In the case of   ideal measurements, 
the functions $u(t)$ and $v(t)$ in Eq.(\ref{fstate}) must always satisfy 
the relation $u(t)^2 + v(t)^2=1$, however this relation may be violated
 in the case of  measurements which introduce significant
level of  disturbance to the system being monitored \cite{thilmeas}.

The survival probability  associated with  $N$ measurements (see Eq.(\ref{fstate}) for a single measurement) performed at
regular intervals $\tau$, $P^N(t)= u(\tau)^{2 N} = \exp(-\gamma(\tau) t)$ where $t=N \tau$.
The effective decay rate, $\gamma(\tau)$ at  
small values of $\tau$ is obtained  using ($\Delta$=2, $\hbar$=1)
 Kofman and Kurizky's formalism \cite{ob}
\begin{equation}
\gamma(\tau)= \int_0^{\infty} d\omega J(\omega) F_{\tau}(\omega-{\Delta \Omega}),
\label{eq:overlap}
\end{equation}
Facchi et. al. \cite{facprl} made comparison of   the effective decay rate $\gamma(\tau)$  to the
 natural decay rate (without measurement), and attributed 
a smaller  $\gamma(\tau)$ than the natural decay rate to
the Zeno effect while a larger $\gamma(\tau)$ (than the natural decay rate)
is linked with the anti-Zeno effect. The transition point occurs at the
intersection of the effective decay rate  and the natural one
at a specific time known as jump time.
Eq.(\ref{eq:overlap})  is dependent on the  convolution of two main functions: (a)
the modulating function $F_{\tau}(\omega-{\Delta \Omega})$ and
(b) spectral density $J(\omega )$. The modulating function 
is given by
\begin{equation}
F_{\tau}(\omega-{\Delta \Omega})=\frac{\tau}{2\pi} {\rm sinc}^2\left[ \frac{(\omega- {\Delta \Omega})\tau}{2}\right]
\end{equation}
and  is associated with measurements at intervals of $\tau$. In the limit
$\tau \rightarrow 0$, $F_{\tau}(\omega-{\Delta \Omega}) \rightarrow 0, \gamma (\tau )\rightarrow 0$ which
explains the emergence of the quantum
Zeno effect that is independent of the  spectral density $J(\omega)$.

\begin{figure}[htp]
  \begin{center}
    \subfigure{\label{fig01}\includegraphics[width=7cm]{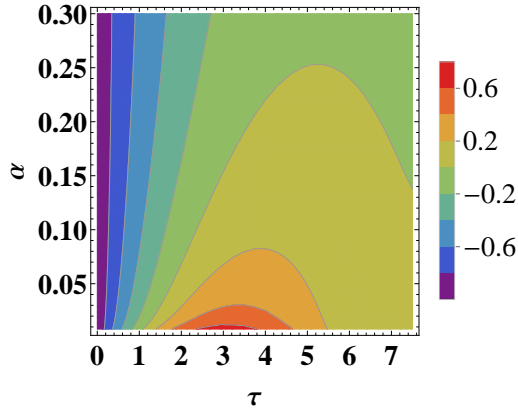}}\vspace{-1.1mm} \hspace{1.1mm}
          \end{center}
  \caption{ 
Occurrence of  the quantum Zeno effect or the anti-Zeno effect,
based on the ratio of decay rates $\frac{\gamma (\tau)}{\gamma _{0}}$,  as a function of  measurement time duration
$\tau $  and $\alpha$, the characteristic frequency of
the bath. The spectral density function, $J(\omega)$ in Eq.(\ref{E2}) is numerically evaluated
using a unit system in which $\hbar$=1, $\Delta =2$ and  $\Delta \Omega$=1. 
The quantum Zeno effect is marked by   regions with negative values  while 
the anti-Zeno effect is denoted by positive values. The occurrence of
these effects  are independent of the coupling strength, $\Lambda$ (using RWA).}
 \label{fig0}
\end{figure}

A third  term, $f(\omega)$  that incorporates rotating and
counter-rotating terms \cite{zheng} introduces an overall shift in the effective decay rate,
without greatly affecting  further analysis of  non-Markovianity with respect to the system parameters.
We thus employ $f(\omega)$=1 (the rotating wave approximation, RWA)  to simplify the numerical analysis.
We obtain the decay rate  at the limit, $\tau \rightarrow \infty $, as
$\gamma _{0}=\gamma (\tau \rightarrow \infty ) \approx J(\Delta \Omega) $.
The  quantum Zeno (anti-Zeno)  effect occurs when $\gamma (\tau ) < \gamma _{0}$
($\gamma (\tau ) > \gamma _{0}$). The parameter range of $\tau$ and 
the $\alpha$, the characteristic frequency of
the bath,  for which these two 
distinct effects occur is shown in Figure~\ref{fig0}. The quantum Zeno effect always
occurs at very short $\tau < 1$, however beyond $\tau > 1$ the 
characteristic frequency of the bath 
appears to influence the occurrence of the anti-Zeno effect  at $\alpha < 0.25$.

While the decay rate itself is dependent on the coupling strength term $\Lambda$,
the occurrence of Zeno or anti-Zeno effect is independent of $\Lambda$, this is partly 
due to the use of the rotating wave approximation. The influence of $\Lambda$ in 
determining Zeno or anti-Zeno dynamics may resurface with removal of the latter approximation,
however  it is expected to be weak compared to the influence of
$\alpha$,  the characteristic frequency of the bath. This may be attributed
to the non-Markovian times-scales at which  Zeno dynamics operate.
The results here (to be shown in Section\ref{joint}) appear to  be influenced strongly
by $\alpha$, which controls the range of  time scales to  which the
bath can respond, and which  is intricately linked to  measurements taken at intervals of $\tau$.
The results in Figure~\ref{fig0} therefore highlight the critical role of $\alpha$ in 
systems where the RWA can be justified. 
Moreover these findings  suggest the relevance of the  Zeno/anti-Zeno effect to inter-qubit non-Markovianian
dynamics rather than intra-qubit non-Markovianity, in structures with large
network size and  connectivity as is the case with 
 the light-harvesting biochemical systems. We therefore 
examine the link between the inter-qubit Non-markovianity and the intra-qubit
Zeno effect in a  qubit pair system, next.

 %%%%%%%%%%%%%%%%%%%%%%%%%%%%%%%%%%%%%%%
%%%%%%%%%%%%%%%%%%%%%%%%%%%%%%%%%%%%%%%%%

\section{Inter-qubit Non-markovinity and the intra-qubit
Zeno effect in a  qubit pair system}\label{joint}

We now extend the  composite system in Eq.(\ref{evolve}) to examine 
 the  joint evolution of a  pair of two-level qubit systems 
in uncorrelated reservoirs via the following  Bell-like initial   state
\begin{eqnarray}
\ket{\Phi}_0 &=& \left[ a\ket{0}_{\mathrm{q1}}\ket{0}_{\mathrm{q2}}
+ b\ket{1}_{\mathrm{q1}}\ket{1}_{\mathrm{q2}} \right]
 \ket{0}_{\mathrm{r1}} \ket{0}_{\mathrm{r2}},
 \label{fstate2}
\end{eqnarray}
where $i$=$1,2$ denote the two qubit-reservoir systems associated
function $u_i(t)$ in  Eq.(\ref{fstate}). 
 $a,b$ are real coefficients and satisfy, $a^2+b^2=1$.
Using  Eq.(\ref{fstate})   and tracing  out the reservoir states we obtain
a  time-dependent qubit-qubit reduced density matrix 
in the basis $(\ket{0 \;0},\ket{0 \;1}\ket{1 \;0}\ket{1 \;1})$ which
evolves with time duration $\tau$ as
\begin{eqnarray}
\label{matrix1}
\rho_{_{\mathrm{q1,q2}}}(t)=
\left(
\begin{array}{cccc}
 f_1& 0 & 0 &f_5 \\
  0 &  f_2 & 0 & 0 \\
  0 & 0 & f_3 & 0 \\
  f_5& 0 & 0 & f_4\\
\end{array}
\right).
\end{eqnarray} 
where $f_1=a^2+b^2v_1(\tau)^2 v_2(\tau)^2$, $f_5=a b  u_1(\tau)\; u_2(\tau)$,
$f_2=b^2 v_1(\tau)^2 u_2(\tau)^2$, $f_3= b^2 u_1(\tau)^2  v_2(\tau)^2$, 
$f_4= b^2 u_1(\tau)^2 u_2(\tau)^2$.
We assume that the usual unit trace
and positivity conditions of the density operator
 $\rho_{_{\mathrm{q1,q2}}}$ are satisfied.
The  reservoir-reservoir  reduced density matrix 
$\rho_{_{\mathrm{r1,r2}}}$ is similarly  obtained by 
by tracing out qubit states. Each  non-zero matrix term of 
$\rho_{_{\mathrm{r1,r2}}}$ is easily obtained from the corresponding 
term $ \rho_{_{\mathrm{q1,q2}}}(t)$  by swapping 
 $u_i \leftrightarrow v_i$. Both  matrices possess the well-known 
 $X$-state structure which preserve its form
during evolution. 
Using  Eq.(\ref{fstate})   and  tracing  out the one qubit and reservoir state, we also obtain
the inter-system qubit-reservoir density matrix $\rho_{_{\mathrm{q1,r2}}}(t)$
of the same form as in  Eq.(\ref{matrix1}), but one in which
$f_1=a^2+b^2u_1(\tau)^2 v_2(\tau)^2$, $f_5=a b  u_1(\tau)\; v_2(\tau)$,
$f_2=b^2 v_1(\tau)^2 v_2(\tau)^2$, $f_3= b^2 u_1(\tau)^2  u_2(\tau)^2$, 
$f_4= b^2 u_1(\tau)^2 v_2(\tau)^2$.

\begin{figure}[htp]
  \begin{center}
    \subfigure{\label{fig2a}\includegraphics[width=3.25cm]{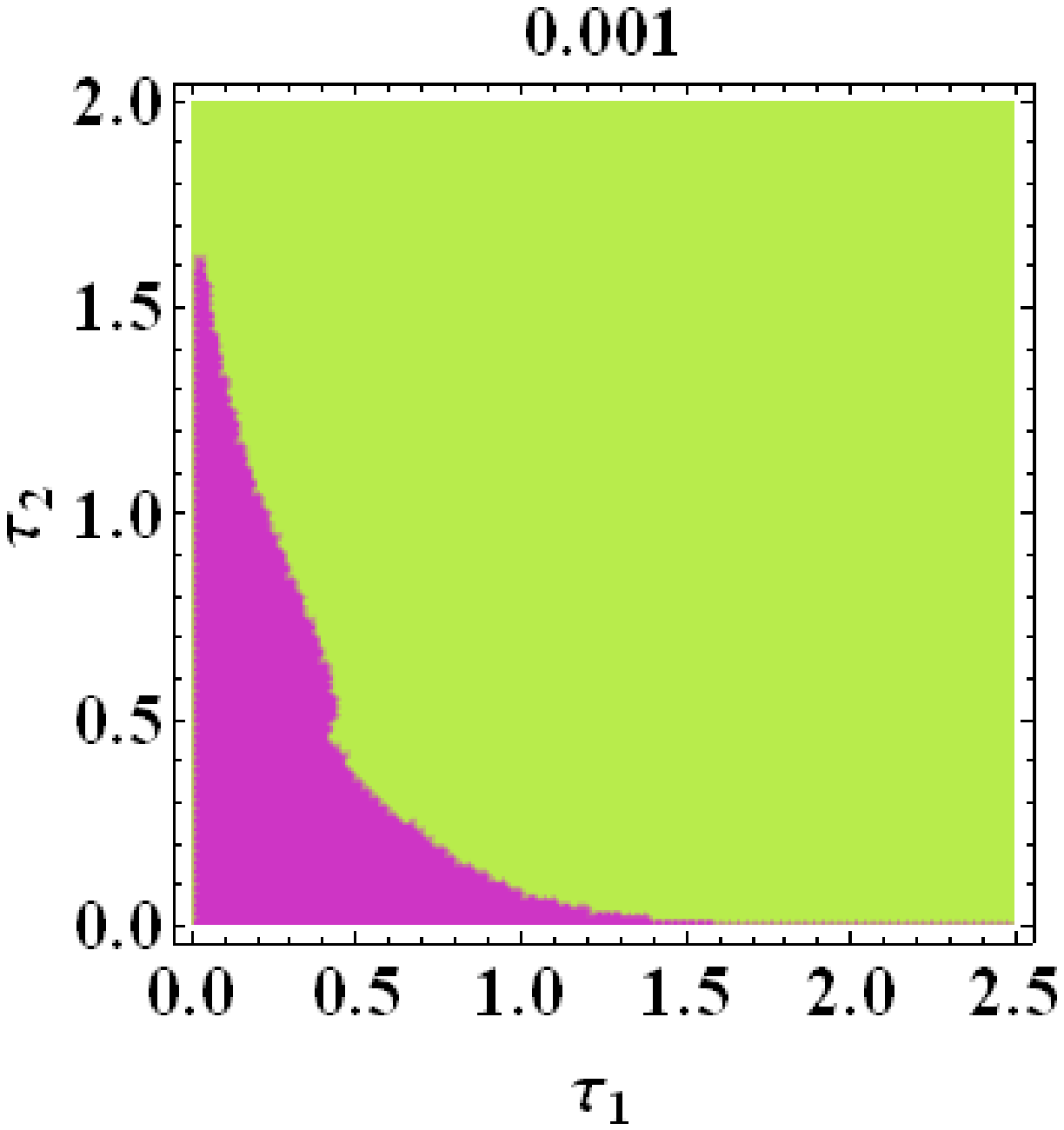}}\vspace{-1.1mm} \hspace{1.1mm}
     \subfigure{\label{fig2b}\includegraphics[width=3.25cm]{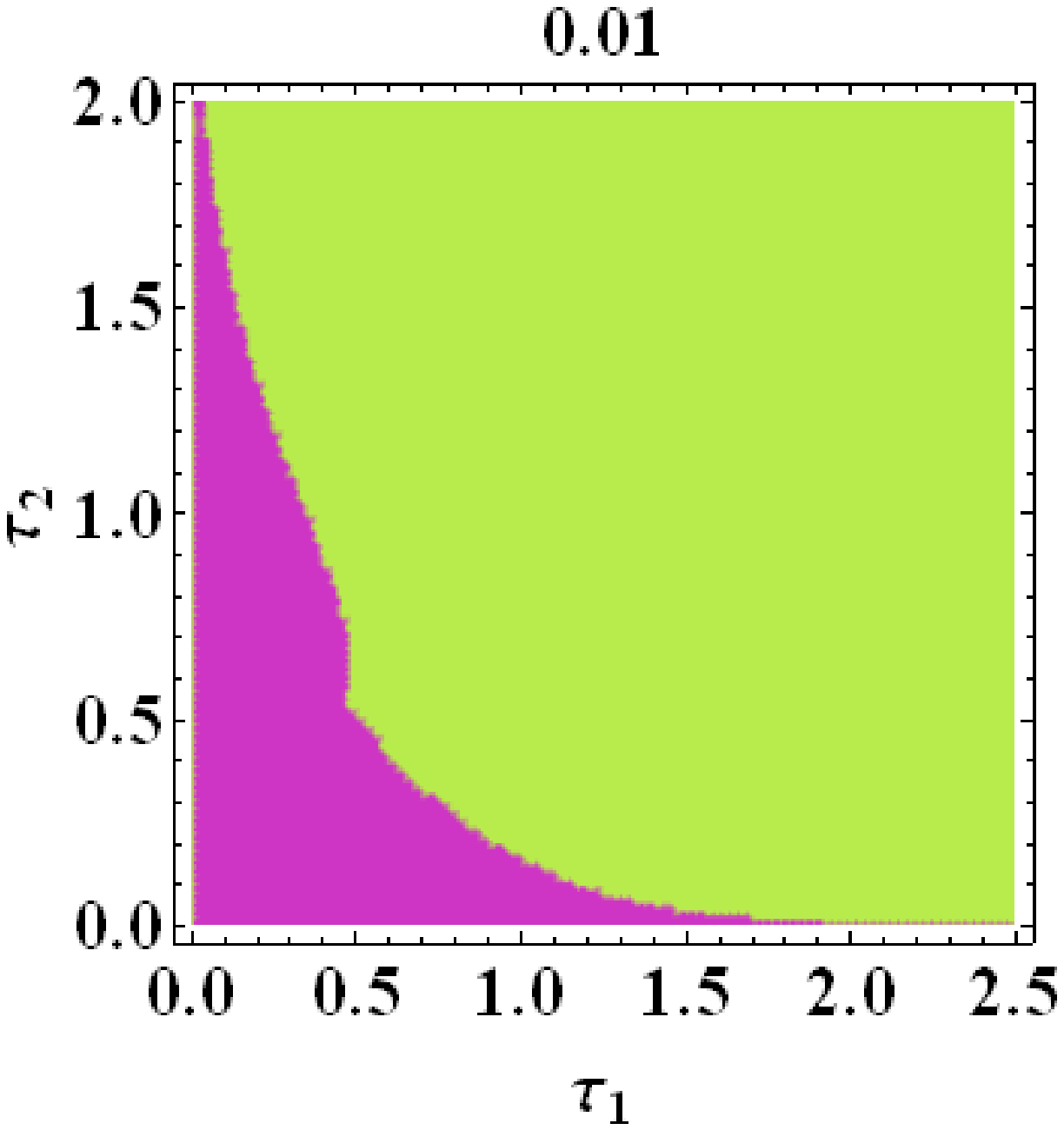}}\vspace{-1.1mm} \hspace{1.1mm}
 % \subfigure{\label{fig3c}\includegraphics[width=3.25cm]{qc}}\vspace{-1.1mm} \hspace{1.1mm}
 \subfigure{\label{fig2d}\includegraphics[width=3.25cm]{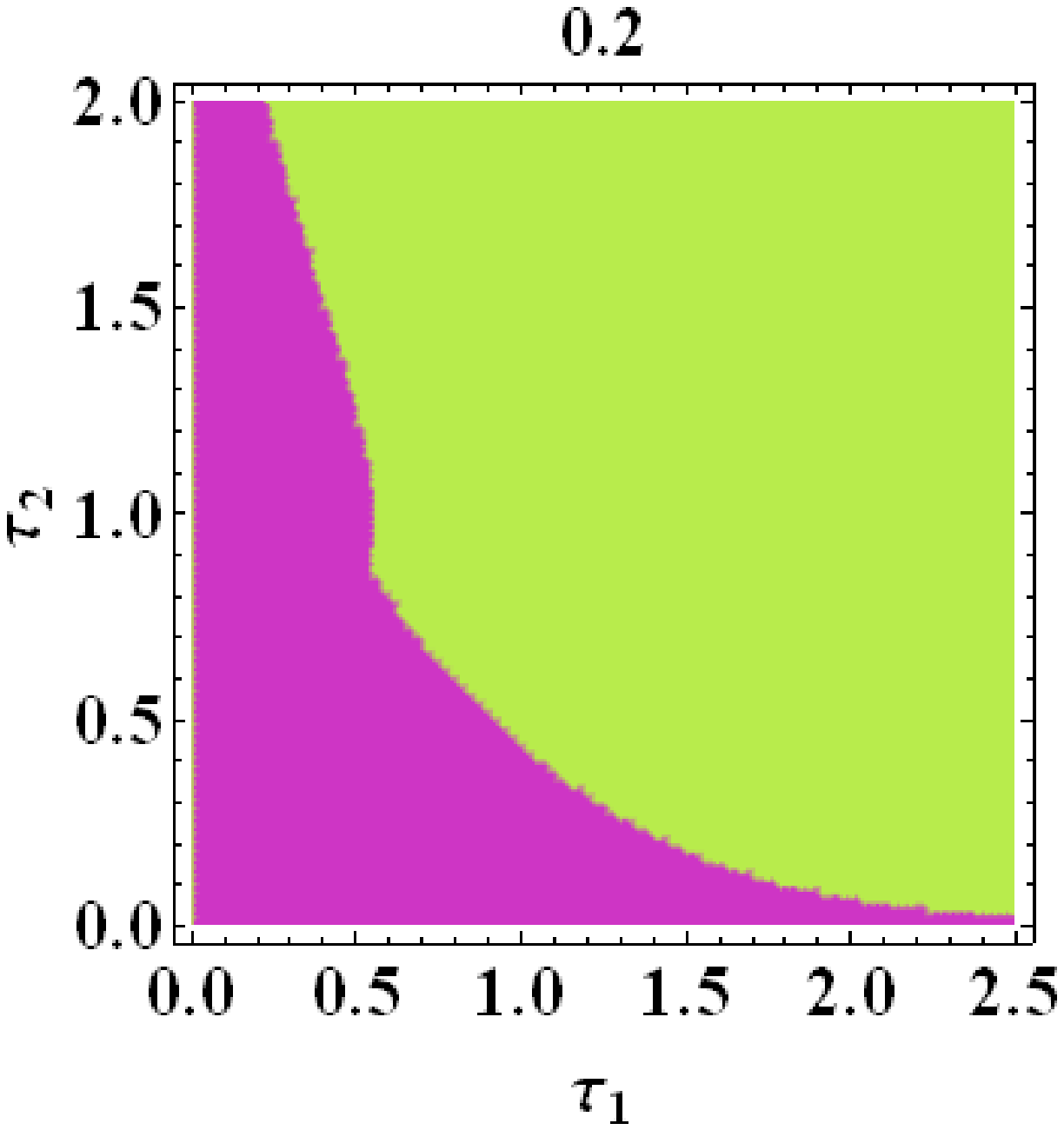}}\vspace{-1.1mm} \hspace{1.1mm}
\subfigure{\label{fig2e}\includegraphics[width=3.25cm]{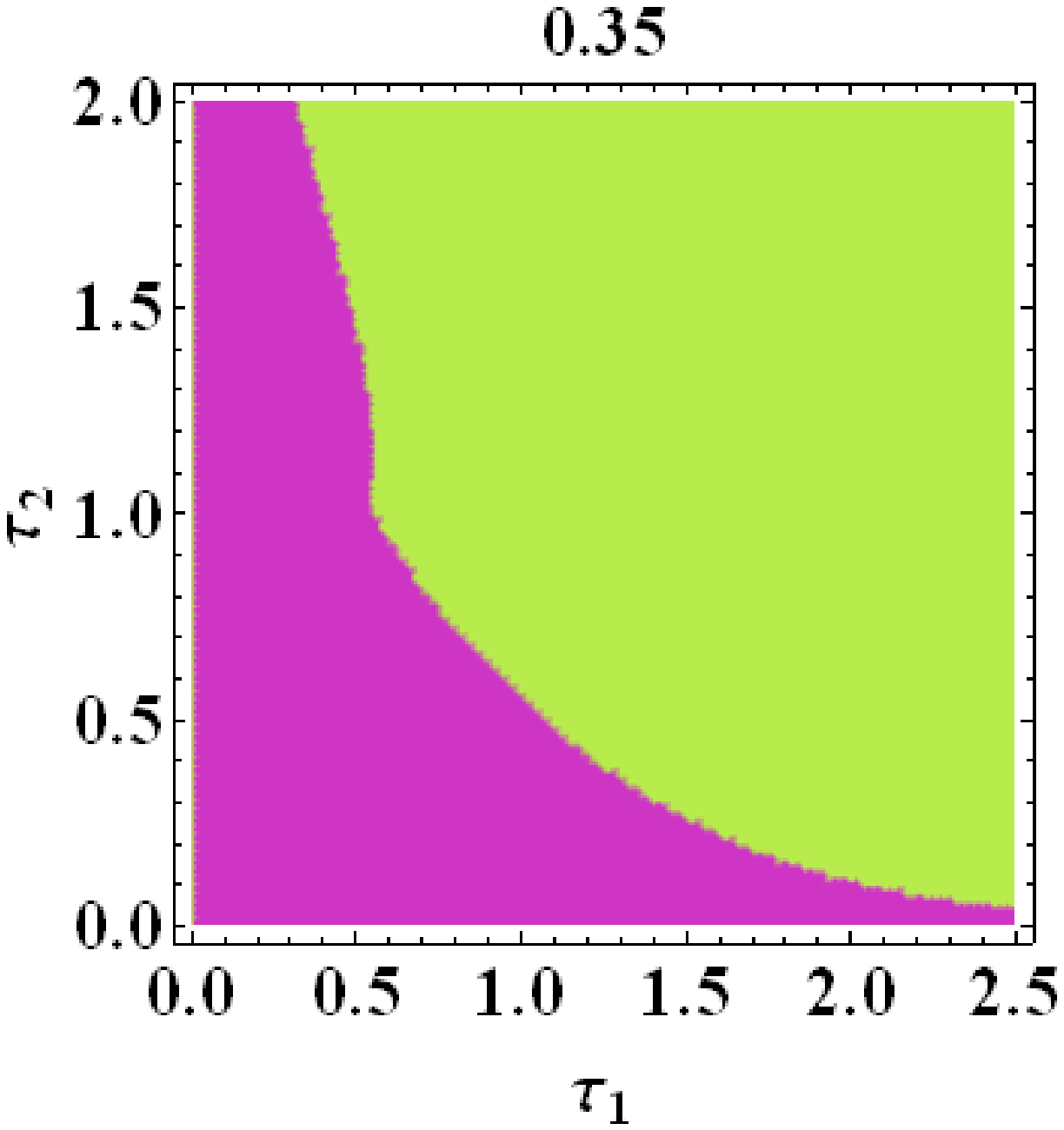}}\vspace{-1.1mm} \hspace{1.1mm}
        \end{center}
  \caption{ Sign of Trace-distance difference function  Sign[$\Delta(\tau_1, \tau_2)$] (see  
 Eq.(\ref{trm})) as a function of $\tau_1$ and $\tau_2$ evaluated for the reduced density matrix corresponding to the qubit-qubit subsystem (Eq.(\ref{matrix1})).
Sign[x] yields  -1, 0, or 1 depending on whether x is negative, zero, or positive. 
Values of  $\alpha$, the characteristic frequency of
the bath is indicated at the top of each figure.
$a$=$b$=$\frac{1}{\sqrt 2}$, coupling strength $\Lambda$ = 0.01, number of measurements $N$=20.
 Regions of   non-Markovianity are shaded purple, while regions which are
shaded green denote occurrence of  Markovian evolution dynamics.
Non-Markovianity increase occurs in conjunction with the Zeno effect for the 
qubit-qubit subsystem. }
 \label{trace1}
\end{figure}

In Figure~\ref{trace1}, we have plotted the trace-distance difference 
function $\Delta(\tau_1, \tau_2)$ (see Eq.(\ref{trm})) as a function of 
$\tau_1$ and $\tau_2$ (using the qubit-qubit density matrix) at  increasing 
values of $\alpha$, the characteristic frequency of the bath.
We recall  that $\tau_1$ is inversely proportional to the measurement frequency of 
one of the initial state, while  $\tau_2$ is the increase in measurement time duration of the  final states.
We note the enhancement of inter-qubit non-Markovianity  with increase in $\alpha$,
(i.e. increased Zeno effect) at  constant coupling strength, $\Lambda$. These results in combination
with the earlier results obtain in Figure~\ref{fig0}, show that the intra-qubit Zeno effect occurs in conjunction
with inter-qubit non-Markovian dynamics. A decrease in decay rate is linked with reservoir feedback
for the qubit-qubit  subsystem.

%%%%%%%%%%%%%%%%%%%%%%%%%%%%%%%%%%%%%%%%%%%%%%%%
%%%%%%%%%%%%%%%%%%%%%%%%%%%%%%%%%%%%%%%%%%%%%%

\begin{figure}[htp]
  \begin{center}
    \subfigure{\label{fig3a}\includegraphics[width=3.25cm]{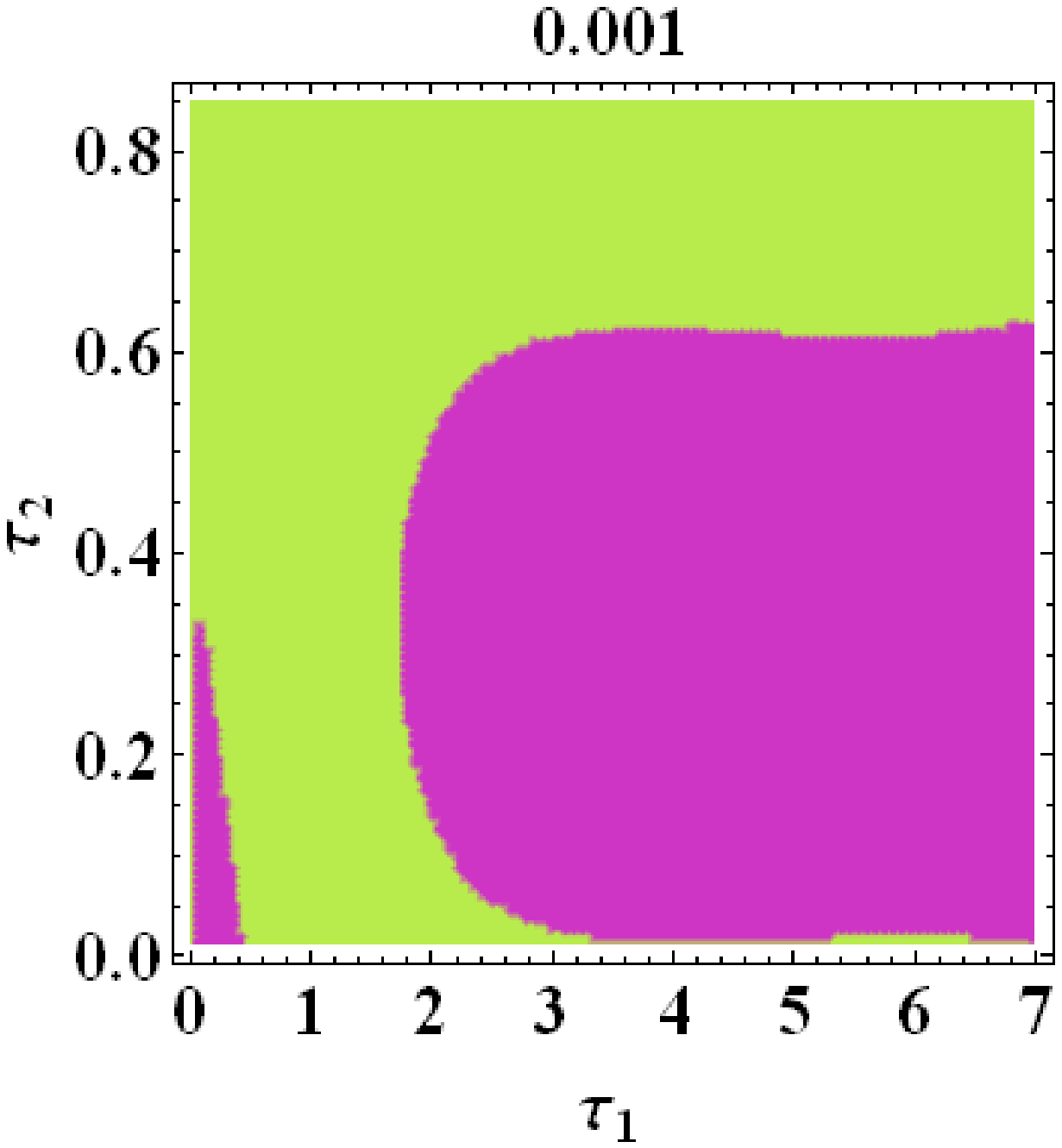}}\vspace{-1.1mm} \hspace{1.1mm}
     \subfigure{\label{fig3b}\includegraphics[width=3.25cm]{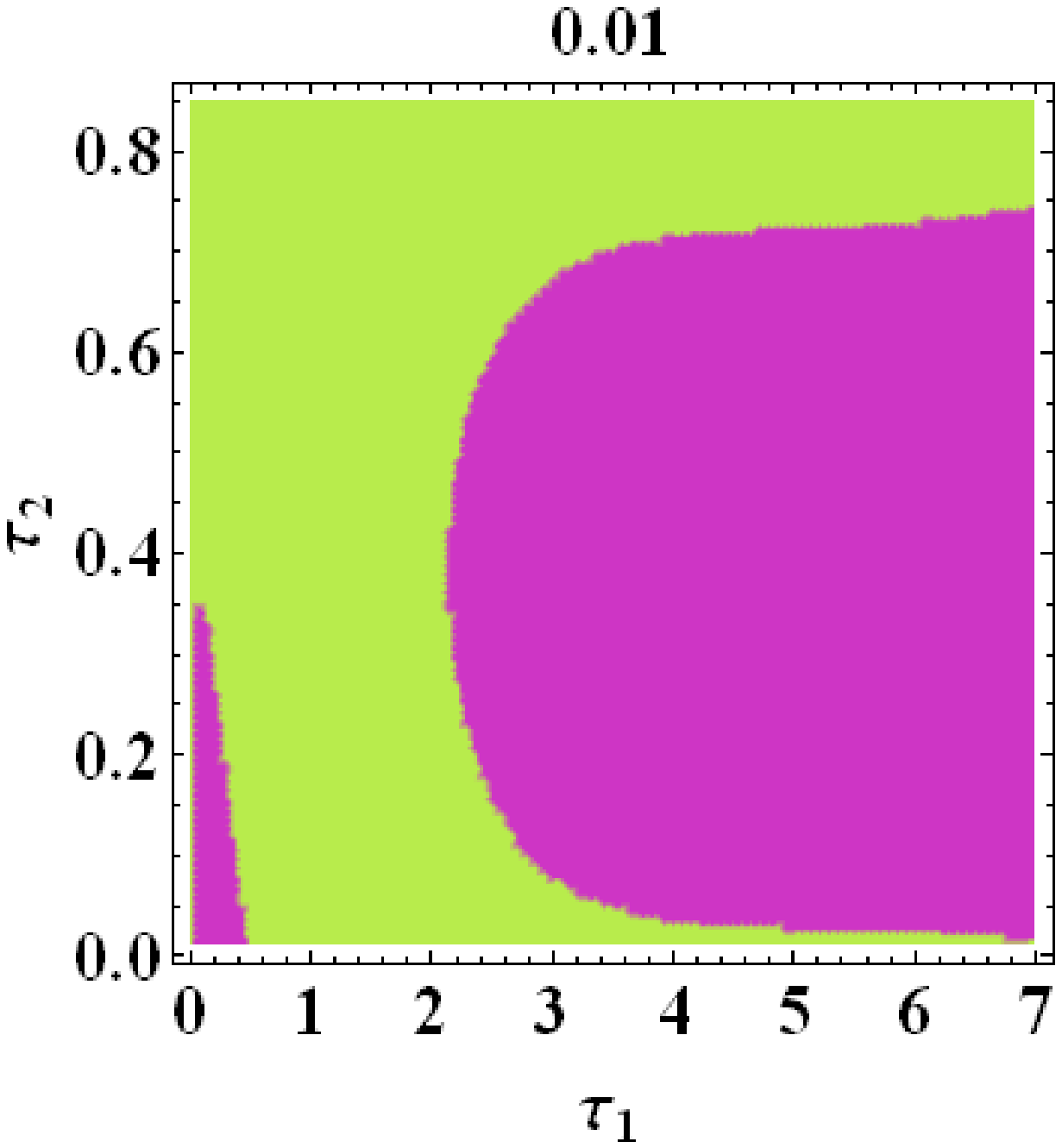}}\vspace{-1.1mm} \hspace{1.1mm}
 % \subfigure{\label{fig3c}\includegraphics[width=3.25cm]{qc}}\vspace{-1.1mm} \hspace{1.1mm}
 \subfigure{\label{fig3d}\includegraphics[width=3.25cm]{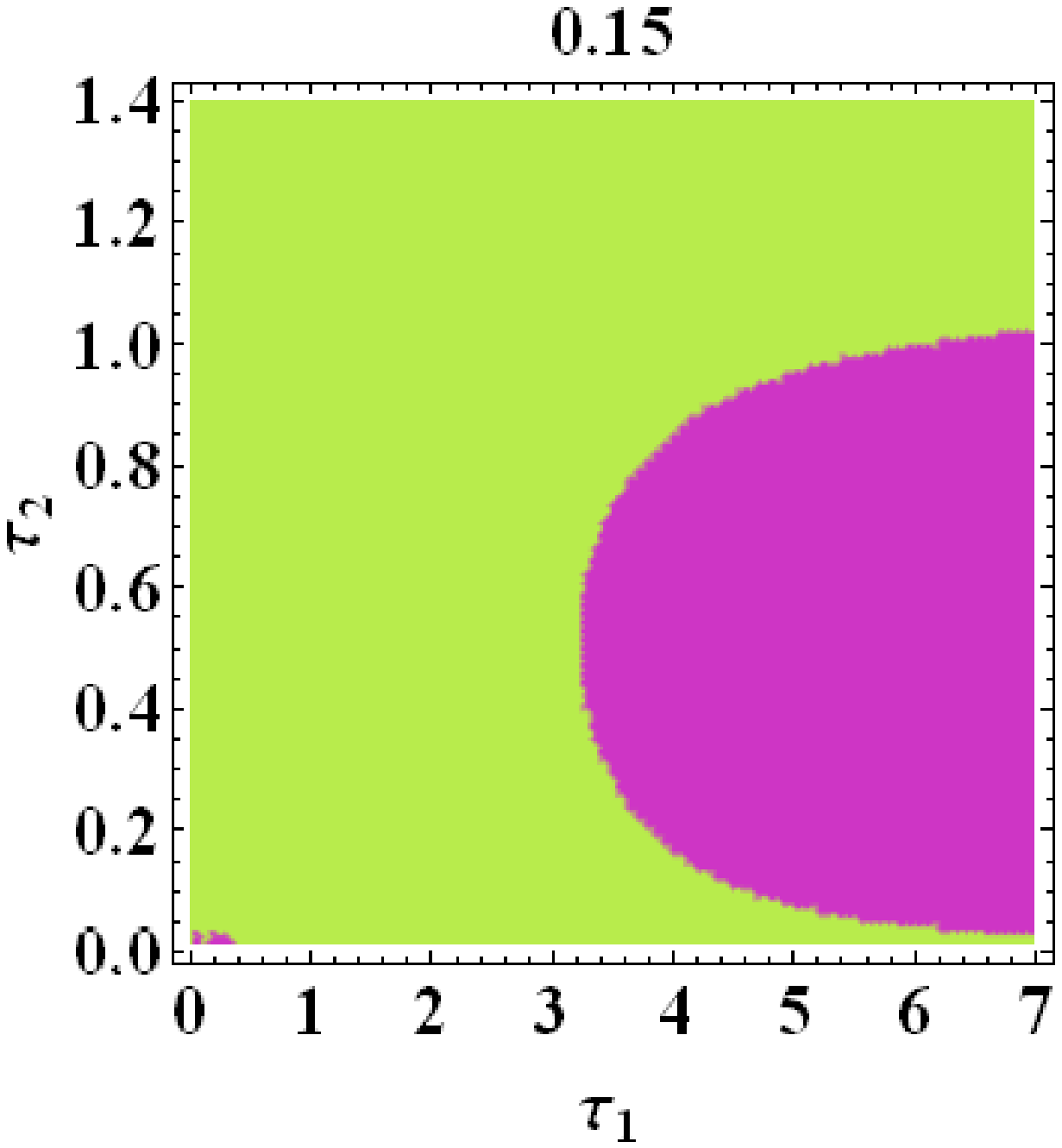}}\vspace{-1.1mm} \hspace{1.1mm}
\subfigure{\label{fig3e}\includegraphics[width=3.25cm]{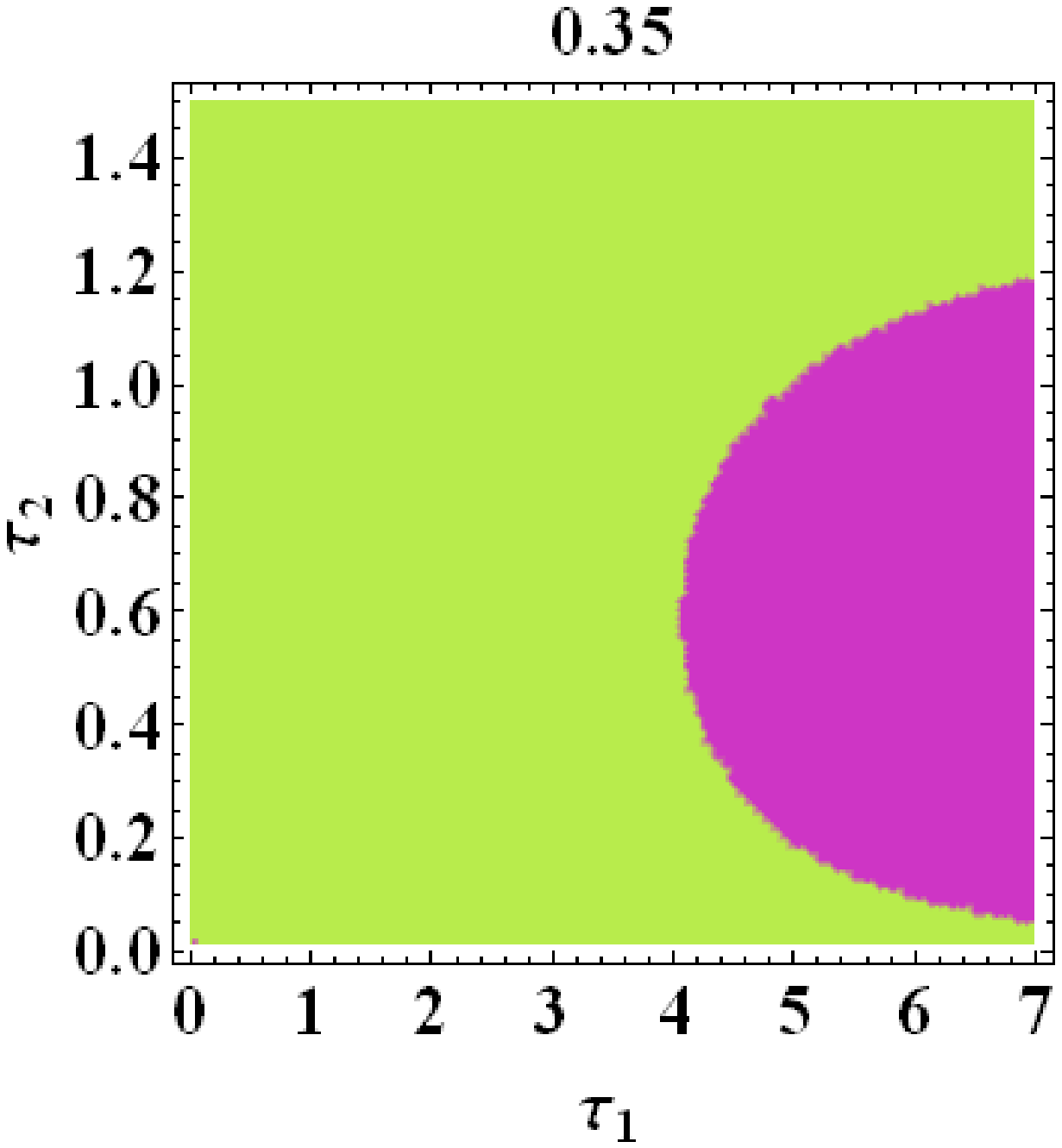}}\vspace{-1.1mm} \hspace{1.1mm}
        \end{center}
  \caption{ Sign of Trace-distance difference function,  Sign[$\Delta(\tau_1, \tau_2)$]  as a function of $\tau_1$ and $\tau_2$ 
evaluated for the reduced density matrix corresponding to the 
inter-system qubit-reservoir density matrix $\rho_{_{\mathrm{q1,r2}}}(t)$.
Values of $\alpha$, the characteristic frequency of
the bath are indicated at the top of each figure.
$a$=$b$=$\frac{1}{\sqrt 2}$, $\Lambda$ = 0.01, number of measurements $N$=20. Regions of 
 non-Markovianity are shaded purple, while regions which are
shaded green denote occurrence of  Markovian evolution dynamics.
Areas of non-Markovianity increases with $\alpha$ at higher $\tau_1 > 2$, while
presence of non-Markovianity is erased for $\tau_1 < 0.1$ for increasing $\alpha$.
 }
 \label{trace2}
\end{figure}

In the case of  the inter-system qubit-reservoir density matrix $\rho_{_{\mathrm{q1,r2}}}(t)$,
 we note that the inter-qubit non-Markovian
dynamics is erased at  $\tau_1 < 0.1$  and enhanced at higher $\tau_1 > 2$  with increase in $\alpha$ 
(see Figure~\ref{trace2}). The results in Figure~\ref{trace2} indicate increasing feedback into
the qubit-reservoir subsystem, consistent with
the intra-qubit anti-Zeno effect that applies for  the specified
range of $\alpha$ shown in Figure~\ref{fig0}. The combined
results of  Figs~\ref{trace1} and \ref{trace2} indicate that 
 decrease (increase) in decay rate due to Zeno (anti-Zeno) effect may  be linked to
information flow between  specific  two-qubit partitions.

While the  results obtained here are specific to the choice of the 
Bell-like initial   state provided in Eq.(\ref{fstate2}), the output 
related to the degree of correlation or anti-correlation of the 
inter-qubit non-Markovinity with respect to the intra-qubit Zeno effect, 
will vary according to the choice of the initial state. The robustness of
the connection between non-markovinity and Zeno effect is based on 
 the area size  of the non-Markovian region (i.e. the sign rather than the magnitude of the 
Trace-distance)   which is evaluated
numerically (coloured purple in Figures~\ref{trace1}, \ref{trace2}). 
The determination of  the size of this region as a function of the amplitude $a$, coupling function $\alpha$, 
and number of measurements $N$  is expected to provide further
information on the non-Markovian dynamics, but will  be  pursued elsewhere.

 We tested the  links
between the Zeno/anti-Zeno effect and inter-qubit non-Markovianity
for all possible two-qubit partitions possible within Eq.(\ref{fstate2}),  and  noted a definite 
dependence on the partition choice (qubit-qubit, reservoir-reservoir, qubit-reservoir)
in relation to non-Markovian dynamics. 
We have focussed
on the two main subsystems which showed significant contrasting modes of  
information flow with respect to changes in decay rate arising from  Zeno or anti-Zeno effect.
The results obtained here show that the bath memory time  (as seen in changes
due to  $\alpha$)  plays a critical role in  the non-Markovian processes of qubit-bath interaction, and
may provide a   basis for the Zeno effect-non-Markovianity link. There is approximate
matching in the time domains involved during effective 
Zeno or anti-Zeno dynamics and the non-Markovian time scale  of the reservoir correlation
dynamics. We bear in mind that these results are obtained on the basis of the RWA and the assumption
of a constant coupling strength, $\Lambda$. 
In the next Section, we consider a tripartite qubit state constructed from a simple model of the
light-harvesting complex, accordingly we adopt slightly different criterias for determining
the Zeno-effect and non-Markovian regions.

%%%%%%%%%%%%%%%%%%%%%%%%%%%%%%%%%%%%%%%%%%%%%%%%
%%%%%%%%%%%%%%%%%%%%%%%%%%%%%%%%%%%%%%%%%%%%%%

\section{Zeno effect  at dissipative sinks in  the photosynthetic reaction center}\label{lhsec}

It is well known \cite{hang,schult,pom} that  several chemical and physics systems can be modeled
using  a reaction coordinate linked to an  effective potential energy
function with two distinct minima points, and which is amenable to analysis
in terms of  exchange of quantum correlations between the system under study 
and its immediate environment. 
The  light harvesting complex is an important system where
the observed long lived coherences, lasting several picoseconds,
 may be  linked to quantum information processing features
inherent in the   propagating  exciton (or correlated electron-hole pair).
The  biological pigment-protein complex, 
called FMO (Fenna-Matthews-Olson) trimer complex  in the sulphur bacteria
\cite{flem} constitutes three symmetrically  equivalent
 monomer subunits. Each monomer unit is generally modeled as a network of
 eight bacteriochlorophyll
(BChl)a molecules within a matrix of protein molecules.  The series 
of  excitonic exchanges between molecular sites is 
described by an exciton  Hamiltonian of the form in  Eq. (\ref{lind}).
Local Lindblad terms  which represent  dephasing and
dissipation processes linked to the surrounding environment
are incorporated during the exchange mechanisms.
The exciton energy is irreversibly transferred to a reaction center (RC), where it 
undergoes  conversion into chemical energy.

Earlier works \cite{caru} examined the fine
interplay between quantum coherence and enviromental noise (both Markovian and non-Markovian)
to achieve optimal  functionality in photosynthetic  systems. 
In a separate study \cite{reben2}, the non-Markovian processes were seen to be 
prominent in the  reorganization energy regime, also noted by   Silbey and coworkers \cite{silbey}
 who showed that the reorganization energy and the bath relaxation rate played critical roles 
during the energy transfer process. Chen et. al. \cite{recom}
presented results which showed the importance of  incorporating 
 the site energy  based electronic coupling correlations.
The  increased oscillations of entanglement   in the non-Markovian
regime of  intra-qubit systems in an earlier work \cite{thilchem2}  further highlighted  the importance of  
feedback mechanisms  in the FMO monomer  system. 
However in all these works, there was
focus only on the intra-qubit Markovian attribute, and to this end, the 
contributions from inter-qubit Markovian dynamics of photosynthetic systems
 need further study.  

An important reason to consider  inter-qubit Markovian dynamics is associated with the  significance 
of  multipartite states  that act as sources of  local and non-local correlations
in large photosynthetic membranes which feature in many FMO complexes \cite{thilchem2}.
The intra-qubit  Markovian (and non-Markovian) model, which is suitable for the 
FMO monomer, is less applicable in the  examination of quantum correlations and 
exchanges  across a wider networked molecular system
with capacity to hold a large number of entangled excitonic qubits.
The latter system may arise when entangled photons 
are absorbed by the photosynthetic membrane, setting up fruitful conditions for
a system of initially correlated chromophores.
In the FMO monomer,  it was shown recently \cite{olb} that  the unique location of the 
eighth chromophore (at site $8$)  gives rise 
to strong inter-monomer interactions, facilitating  
excitation transfer between  monomers of the FMO trimer.
Hence our approach focussed on inter-qubit non-Markovian dynamics
is best suited  to the  realistic condition  of an entire photosynthetic 
membrane constituting many FMO complexes and thousands of bacteriochorophylls.
In Section \ref{joint}, we noted the time domains involved during effective 
Zeno mechanism appears matched with the non-Markovian time scale  
of the reservoir correlation dynamics. This points to the relevance of the
 inter-qubit Markovian dynamics  in the context of the Zeno mechanism 
in applications related to light harvesting systems.

Here we utilize the tripartite states arising from 
a donor-acceptor-sink model to examine the subtle link between
the intra-qubit Zeno effect and the inter-qubit non-Markovian dynamics.
The Zeno criteria used is dependent on the indirect action of the dissipative  sink
on the  non-Markovian dynamics of a specific inter-qubit system. The
non-Markovianity is evaluated using the 
trace-distance difference measure of a  tripartite state, these
criterias therefore differ from those used in 
Section \ref{spinbo}. A  prototypical  energy transfer mechanism in 
light-harvesting  systems  is thus modeled via a 
photosynthetic reaction center (RC) consisting of the 
donor and acceptor protein pigment complexes  and a third site
 acting as the phonon-dissipative sink with 
 a continuous frequency spectrum. The energy from the acceptor is considered
to  dissipate into the phonon  reservoir over  time. The 
 effective Hamiltonian, $\hat{H}_T$,  which  describes  of the donor and acceptor coupled
to the sink consists of  an exciton at the donor site $d$,   coupled to second exciton 
at the  acceptor $a$ site, which in turn is   interacting with its own source of bosonic reservoir $r$  ($\hbar$=1)
 \bea
\label{ham}
    \hat{H}_T&=&\hat{H}_s+\hat{H}_r+\hat{H}_I\,  \\
   \hat{H}_s &=&  \omega_d \; \sigma^d_+\; \sigma^d_-+\omega_a \; \sigma^a_+\; \sigma^a_- + 
V (\sigma^d_-\, \sigma^a_+ +  \sigma^d_+\, \sigma^a_-)\, 
\label{hamS} \\
    \hat{H_r} &=& \sum_k \omega_k b^\dag_k b_k, \quad 
    \hat{H_I} =\sum_{k=1}^{N}(\varphi_{k}\sigma^a_- \; \hat{b}_{k}^{\dag}+ \varphi_{k}^{*}\hat{b}_{k} \; \sigma^a_+),
\label{hamR}
\eea
where  $\omega_d$ ($\omega_a$) is the exciton resonance frequency at the donor (acceptor) site , 
$\sigma^d_+ ({\sigma^d}_-)$ denotes the raising (lowering) operator of the  exciton \cite{thilopera} at the donor site,  
and $\sigma^a_+ ({\sigma^a}_-)$ denotes the raising (lowering) operator of an  exciton at the acceptor site.
 The operator $\hat{b}_{k} \; (\hat{b}^{\dag }_{k})$ annihilates (creates) a phonon with frequency $\omega_k$ in $k$-th mode  of the reservoir. $V$ is the coupling constant between the acceptor and donor excitons 
 and  $\varphi_{k}$ is the linear coupling 
between the exciton  at the acceptor site and the phonon  with frequency $\omega_k$ in $k$-th mode.

Using Feshbach's projection-operator method \cite{fesh},  
 the total Hilbert space of  $ \hat{H}_T$ is  divided into two orthogonal subspaces 
generated by  the projection operators,
${\cal P}$ and ${\cal Q}$ where  ${\cal Q}=1-{\cal P}$. For the 
photosynthetic reaction center (RC), ${\cal P} =  \sigma^d_+\; \sigma^d_-
+  \sigma^a_+\; \sigma^a_-$ and
${\cal Q}= \sum_k  b^\dag_k b_k$, such that ${\cal P Q}= {\cal Q P}$=0.
The density operator associated with the qubit-cavity system  is  given by $\rho_s(t)={\cal P} \; \rho {\cal P}$,
$\rho$ being  the density operator of the total system.
Using a phenomenological master equation that incorporates only energy dissipation 
from the exciton at the acceptor site  (with decay rate $\lambda_c$) , and providing for 
the presence of a  reservoir system with a flat spectrum,
the reduced density matrix, $ \rho_s$ is obtained  as
\bea
\label{pme2}
	\frac{\mathrm{d}}{\mathrm{d}t} \rho_s &=& i( \rho_s \; H'_s - H'_s \; \rho_s) +
 \lambda_c \sigma^a_-  \rho_s \sigma^a_+ , \\ \label{pme3}
H'_s &=&
	\omega_d \; \sigma^d_+ \;  \sigma^d_- + (\omega_a + \; \delta - \; i \frac{\lambda_c}{2}) \sigma^a_+ \sigma^a_-
+ V (\sigma^d_-\, \sigma^a_+ +  \sigma^d_+\, \sigma^a_-)\, 
\eea
where $\delta$ arises due to  energy renormalization,  and  the decay rate $\lambda_c$ 
is influenced by the environmental bath parameters such as temperature and
distribution profile of vibrational frequencies. Eq. (\ref{pme2}) is 
 equivalent to the Haken-Strobl model at infinite temperature, 
where pure dephasing is accounted for in terms of a classical, fluctuating field in the 
presence of a sink \cite{silbey}. 
By setting the resonant condition  $ \omega_d = \omega_a +  \delta$, and assuming
the presence of the initial excitation to be  in the donor exciton state,
the multipartite state  of the donor-acceptor-reservoir can be expressed as
\be
 \ket{\psi_t}= \ket{1}_d\otimes\ket{0}_a\otimes\ket{\mathbf{0}}_r,
\label{inist}
\end{equation}
The presence of an exciton at  the donor (acceptor) site 
is denoted by $\ket{1}_d$  ($\ket{1}_a$), whilst  $\ket{0}_d$  ($\ket{0}_a$)
denotes the absence of the exciton at the donor (acceptor) site. 
 The reservoir state $\ket{\mathbf{0}}_r$=$\prod_{k}\ket{0}_{k}$
denotes the vacuum state,  while  $\ket{\mathbf{1}}_r$
is a   collective state consisting of singly  excited states of the form,  $\ket{\mathbf{1}}_{r_i}$=
$\frac{1}{R_0} \sum_{k}\;\varphi_{k} |1_{k}\rangle_{r}$ where $R_0 = 
\sqrt{\sum_k |\varphi_{k}|^2}$, and  states which are 
orthogonal to it \cite{ushthil}.

 On the basis of the  quantum trajectory approach \cite{traj1},
dissipative  processes result in the transfer of states  from the 
exciton donor $d$- acceptor $a$  subspace to the reservoir subspace.
Eq.~(\ref{inist}) evolves under the action of the Hamiltonian  in Eq.~(\ref{ham}) as 
\be
\label{evolve2}
    \ket{\psi_t}=\xi_t \ket{1}_d\ket{0}_a \ket{\mathbf{0}}_r+
\eta_t \ket{0}_d \ket{1}_a \ket{\mathbf{0}}_r + \chi_t \ket{0}_d \ket{0}_a \ket{\mathbf{1}}_r, 
\ee
where $|\xi_t|^2$ ($|\eta_t|^2$) is the probability that the exciton  is present at the
donor (acceptor) site, and  $|\chi_t|^2=1-|\xi_t|^2-|\eta_t|^2$  
 is the probability that the collective reservoir state is excited.
For the initial condition, $\xi_0$=1,  $\eta_0$=0,  and we 
obtain the analytical expressions, 
$|\xi_t|^2= e^{-\lambda_c t/2} \left[\cos{\Omega} t +  \frac{\lambda_c}{4 \Omega}\sin{\Omega} t \right]^2$,
and $|\eta_t|^2 = e^{-\lambda_c t/2} \frac{V^2}{\Omega^2}\sin^2{\Omega} t$,
where the Rabi frequency $2 \Omega= (4 V^2 -(\frac{\lambda_c}{2})^2)^{1/2}$.

%%%%%%%%%%%%%%%%%%%%%%%%%%%%%%%%%%%%%%%%%%%%%%%%
\begin{figure}[htp]
  \begin{center}
    \subfigure{\label{fig4a}\includegraphics[width=5.25cm]{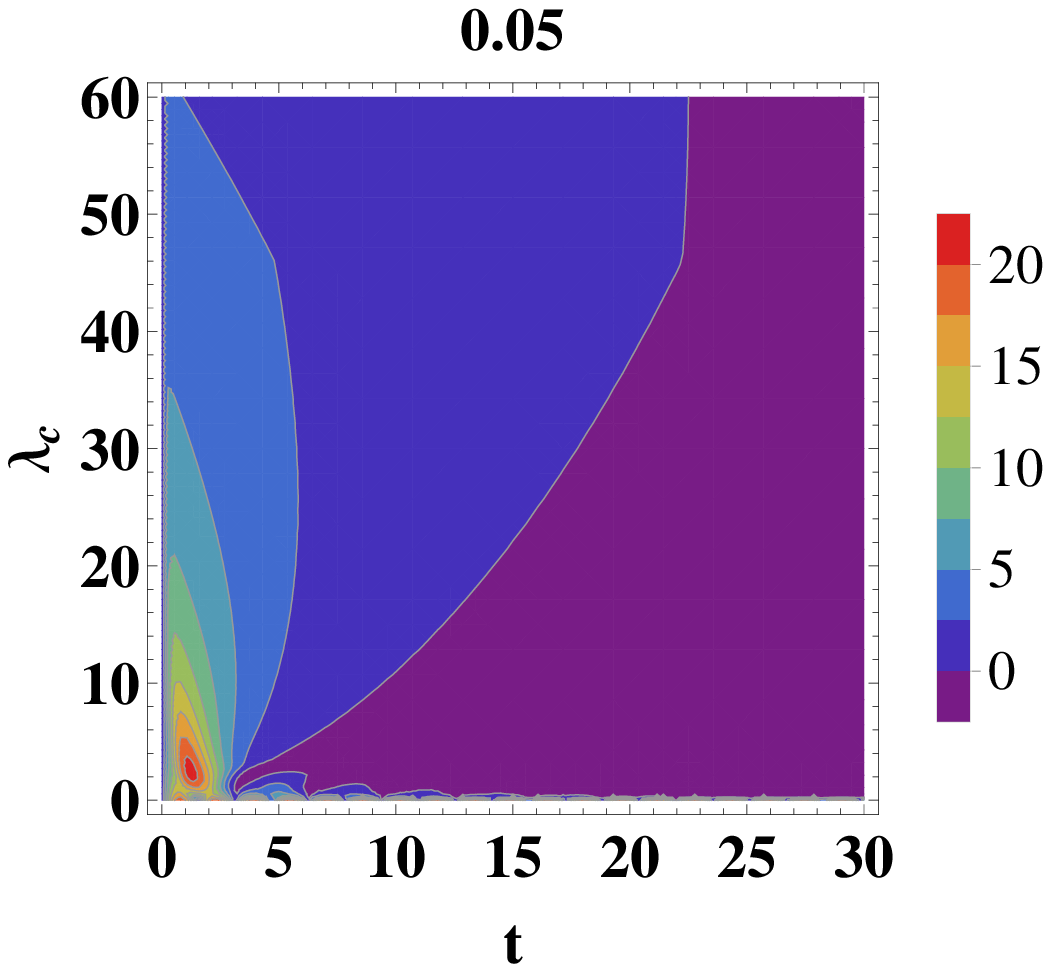}}\vspace{-1.1mm} \hspace{1.1mm}
 \subfigure{\label{fig4b}\includegraphics[width=5.25cm]{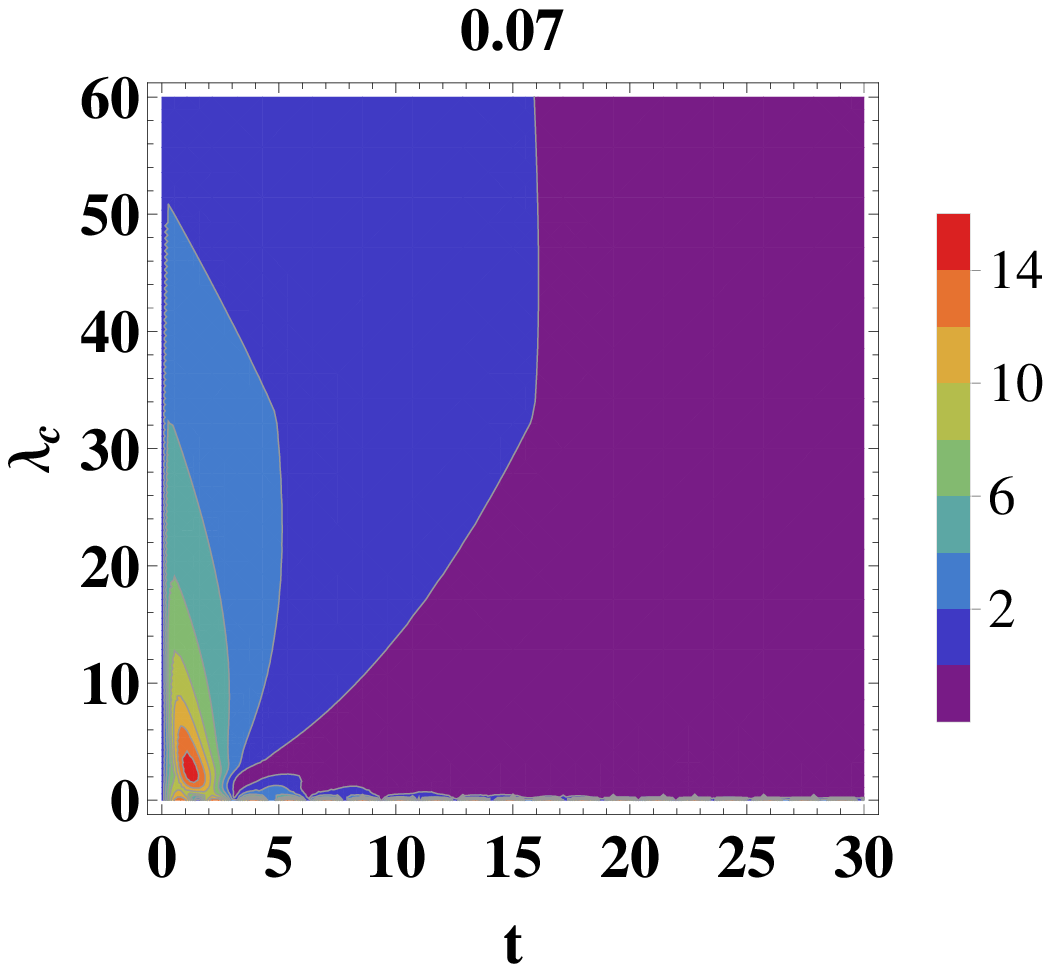}}\vspace{-1.1mm} \hspace{1.1mm}
 \subfigure{\label{fig4c}\includegraphics[width=5.25cm]{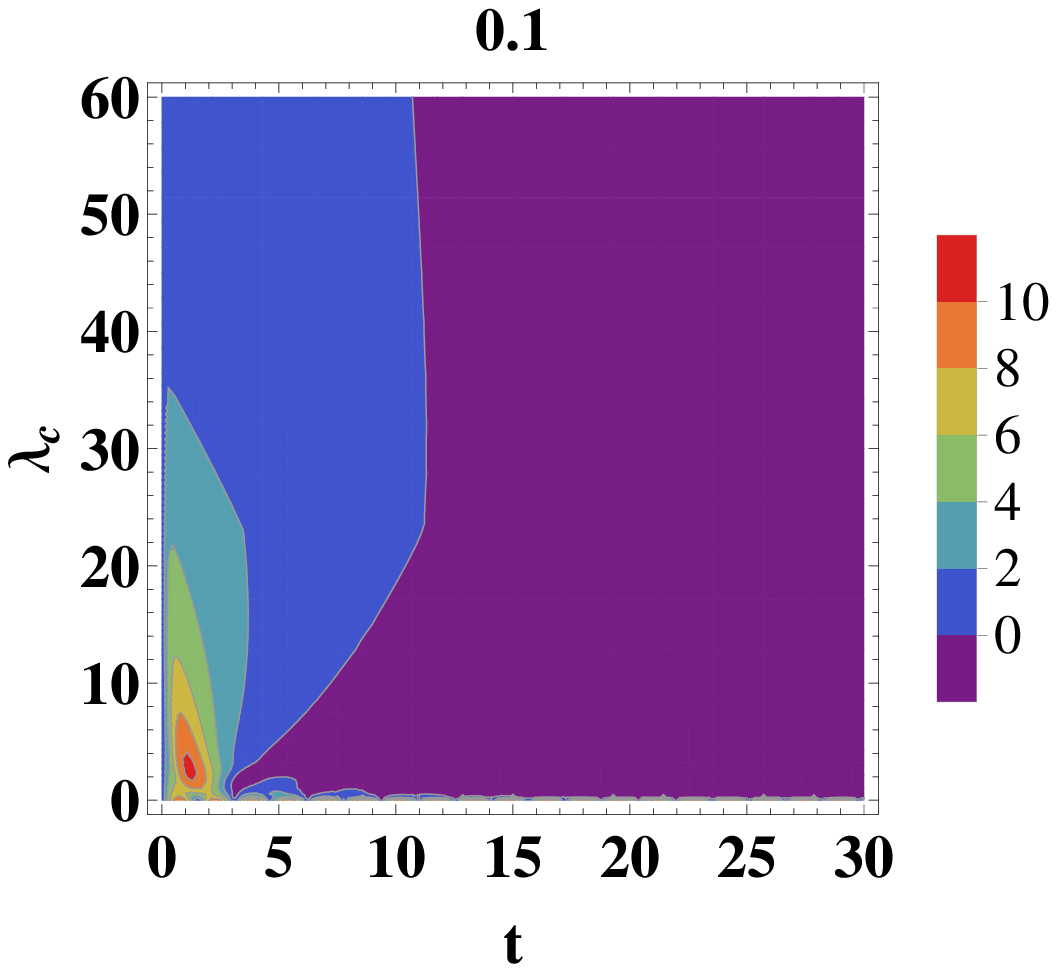}}\vspace{-1.1mm} \hspace{1.1mm}
            \end{center}
  \caption{ Non-Markovianity in the entangled photosynthetic reaction center (RC) for a state of the 
form given in Eq.~(\ref{fstatep1}).   Trace-distance difference function,  
$D(t,\tau)$ (see   Eq.(\ref{trid}))  is evaluated as a function of $t$ and 
decay rate, $\lambda_c$ for the density matrix corresponding to the exciton-donor(1)-sink(1)-sink(2) 
subsystem (Eq.(\ref{tqubitm})). $a$=$b$=$\frac{1}{\sqrt 2}$ and 
values of $\tau$ are indicated at the top of each figure.
Large damping ($\lambda_c$) associated with higher rate of detection by the  dissipative sink, 
appear to  enhance non-Markovian dynamics (positive values of $D(t,\tau)$).
 The unit system  is based on $\hbar=V=$1, with time $t$  obtained as inverse of 
 $\Omega_0$ (at $\lambda_c$ = 0)} 
 \label{lhs}
\end{figure}
%%%%%%%%%%%%%%%%%%%%%%%%%%%%%%%%%%%%%%%%%%%%%%

 Eq.(\ref{evolve2}) can be extended to examine
 the  joint evolution of a  pair of qubit systems (denoted)
in uncorrelated reservoirs via a initial   state of the form
\begin{eqnarray}
\ket{\Phi}_l &=& a  \ket{\psi_t}_1  \ket{\psi_0}_2 + b  \ket{\psi_t}_2  \ket{\psi_0}_1
 \label{fstatep1}
\\
 \ket{\psi_0}_1 &=&  \ket{1}_{d1}\ket{0}_{a1} \ket{\mathbf{0}}_{r1}, \quad \quad
\ket{\psi_0}_2 = \ket{1}_{d2}\ket{0}_{a2} \ket{\mathbf{0}}_{r2}
 \label{fstatep2}
\end{eqnarray}
where  $i$=$1,2$ denote the two distinct qubits and $a,b$ are real coefficients (as in
 Eq.(\ref{fstate2})).  Prominent
non-Markovian features  are seen (Figure~\ref{lhs}) in the tripartite qubit 
associated with  the donor exciton(1)-sink(1)-sink(2) partition:
\be
\label{tqubitm}
 {\rho}_{d_1,r_1,r_2}(t)=\left(
\begin{array}{llllllll}
 \left(|\eta_t| ^2+|\xi_t| ^2\right) b^2+a^2 |\eta_t| ^2 & 0 & 0 & 0 & 0 & 0 & 0 & 0 \\
 0 & b^2 |\chi_t|^2 & a b |\chi_t|^2 & 0 & a b \sqrt{|\xi_t| ^2 |\chi_t|^2} & 0 & 0 & 0 \\
 0 & a b|\chi_t|^2 & a^2|\chi_t|^2 & 0 & a^2 \sqrt{|\xi_t| ^2 |\chi_t|^2} & 0 & 0 & 0 \\
 0 & 0 & 0 & 0 & 0 & 0 & 0 & 0 \\
 0 & a b \sqrt{|\xi_t| ^2 |\chi_t|^2} & a^2 \sqrt{|\xi_t| ^2 |\chi_t|^2} & 0 & a^2 |\xi_t| ^2 & 0 & 0
   & 0 \\
 0 & 0 & 0 & 0 & 0 & 0 & 0 & 0 \\
 0 & 0 & 0 & 0 & 0 & 0 & 0 & 0 \\
 0 & 0 & 0 & 0 & 0 & 0 & 0 & 0
\end{array}
\right)\\
\ee
We employed the   trace-distance difference 
\be
D(t, \tau) = \frac{D[\rho(t),\rho(t+\tau)]- D[\rho(0),\rho(\tau)}{D[\rho(0),\rho(\tau)]}
\label{trid}
\ee
to identify violations of the  monotonically  contractive characteristic features
in the tripartite state  of Eq.(\ref{evolve}).  
The results in Figure~\ref{lhs} show that a large damping or decay rate ($\lambda_c$)
 associated with the trapping process at the dissipative sink leads to an   
enhancement of non-Markovian dynamics  in the donor exciton(1)-sink(1)-sink(2) partition,
 ${\rho}_{d_1,r_1,r_2}(t)$ (Eq.(\ref{tqubitm})).
The non-Markovian effects is pronounced at small values of $\tau$.
The two dissipative sinks which act as indirect  detectors \cite{ezeno},
appear to induce a anti-Zeno-like effect facilitating 
information feedback into the specific partition considered here.
This result is consistent with those in a recent work \cite{fujii} where it was shown that 
 repeated measurements in disordered systems can  induce a quantum anti-
Zeno effect which enhance quantum transport  under certain conditions.
It is important to note that the  Zeno criteria used here is
 reliant on the detecting role of the dissipative sinks,
which differs from that used in Section \ref{spinbo}
in which a quantitative relation (Eq.(\ref{eq:overlap}))  was employed.

The observation of increased inter-qubit non-Markovian dynamics due
to the dissipative sinks in the protypical donor-acceptor-reservoir
model may be applied to the larger
networked systems of   photosynthetic biomolecules.
A quantitative assessment of the contribution from the
anti-Zeno-like action of the photosynthetic sink 
 to the efficiency of energy transfer is expected to be numerically intense
involving  matrix sizes  larger than the tripartite state  in Eq.(\ref{tqubitm})
Such calculations would  need to take 
into account  extrinsic factors such
as network size and topological connectivity present in
the  molecular structures of multichromophoric macromolecule(MCMM) systems. 

As discussed in Section \ref{basic},
the density matrix in any physically realistic 
subspace, in any   initially correlated system-bath state,  appear not to
necessitate a statistical interpretation. In general, it is desired
to employ exact and non-perturbative methods to examine general non-Markovian dynamics 
of quantum systems. However
 the quantum master equation approach ((Eqs.(\ref{ham}),(\ref{hamS})))  and
subsequent evaluation of the  density matrix of  high dimensional systems  
of large biomolecular system present expected insurmountable challenges.
In this regard, 
the results of the set of states for which non-positivity appears,
should be compared with alternative approaches such as
 the non-perturbative hierarchical equations of motion (HEOM) technique \cite{tani,kreis},
with availability of appropriate  computational resources in  future works.
It is hoped that the donor-acceptor-sink  model used here, will
serve as  a  starting point to comprehend the mechanisms under which
non-Markovian  and dissipative dynamics appear to coexist,
with implications for salient properties of chemical and
biological systems.

\section{Discussion and Conclusions}\label{con}

The observations of long-lasting coherence at ambient temperatures \cite{engel} 
in large biomolecular system
have yet to be satisfactorily explained  due to the sheer complexity
of modeling large complex molecules against a background of noisy processes.
Till now, the exact link between the robust coherence times and 
the time taken for  excitation energy to be
transferred  from the antenna to the reaction center
in the FMO trimer complex remains unknown. This work is a step
forward in a qualitative examination  of possible coordination  between the observed coherence
and the environment-assisted transfer mechanisms based on the Zeno-effect.

Further insight into  understanding of the  mechanisms in natural
systems will provide innovative  ways for robust control 
in open quantum dissipative systems. Photosynthetic complexes have  been modeled as 
quantum channels \cite{noiseC} in Ref.\cite{noise},
where  noise-assisted energy transport have been examined
in  terms of transmission of classical and quantum information,
via characterization by population damping and decoherence 
parameters. The classical and quantum channel capacities of photosynthetic complexes
were seen to be enhanced by the presence of noise unlike the 
noiseless system with zero capacity  limit. 

The role of a  noisy environment which acts as a detection medium with
Zeno-like effects, could be further expanded for applications related to 
quantum control and selection of quantum channels with  capacities of a quantifiable amount.
The result in Figure~\ref{lhs}, even if it is obtained for one specific partition, highlights
the   possibility of quantum control of energy transfer mechanisms through
concatenated sequences and  pulse interval optimization associated with
dynamical coupling theory \cite{dye1,dye2}, using
 artificial light harvesting systems.
It is known that open-loop quantum optimal control techniques \cite{control} 
may be utilized to verify  quantum coherent
transport mechanisms in light-harvesting 
complexes. To this end,   Zeno-like effects may be incorporated in
 optimally shaped laser pulses \cite{control} 
to initialize a photosystem, either in  localized or extended mode,
to discriminate different routes of transport, and to  direct energy along critical pathways
 in   light harvesting complexes. This possibility needs detailed study
to seek effective quantum control for viable applications in energy systems,
and especially in the regime of optimal energy transfer.

In summary, using an entangled system consisting of two  spin-boson and 
and reservoir systems , we have identified enhanced non-Markovian signatures 
which can be associated with the  Zeno or anti-Zeno effect depending on the
subsystem under consideration. The results indicate
that  changes in decay rate due to the Zeno or anti-Zeno effect
 may be  linked to non-Markovian type feedback from the system or reservoir.
The time domains involved during effective 
Zeno or anti-Zeno dynamics is of the same order of magnitude as the non-Markovian time scale  
associated with the reservoir correlation dynamics. 
This study highlights that the context of non-Markovianity
 can reveal  a different phenomenological perspective of the quantum Zeno/anti-Zeno
effects.  Consideration of the Zeno effect in a model photosynthetic system
highlights enhanced inter-qubit non-Markovianity in the tripartite subsystem
involving two dissipative sinks. Further investigation is needed
to seek a rigorous  link between the
efficient rate of energy transfer in light-harvesting
systems and the  mechanism by which the monitoring action of the dissipative sinks  results in 
notable non-Markovian signatures in specific partitions of 
multipartite states. The latter states  hold much relevance in 
  large photosynthetic membranes which constitute  many FMO complexes \cite{thilchem2},
with capacity to hold a large number of entangled excitonic qubits,
when  conditions for occurrence of 
initially correlated chromophores become favourable.

\section{Acknowledgments}

 A. T. gratefully acknowledges the  support of  the Julian Schwinger Foundation Grant,
JSF-12-06-0000.  The author would like to thank F. Caruso and M. B. Plenio
for  useful comments on the initial draft of this manuscript, and the anonymous referees
for helpful suggestions.

\end{document}